\documentclass{aa}

\usepackage{graphicx}
\usepackage{txfonts}
\usepackage{adjustbox}
\usepackage{xcolor}
\usepackage{hyperref}
\hypersetup{colorlinks=true,allcolors=[rgb]{0,0,0.5}}
\usepackage[version=4]{mhchem}

\makeatletter
\renewcommand*\aa@pageof{, page \thepage{} of \pageref*{LastPage}}
\makeatother

\begin{document}

   \title{Fully time-dependent cloud formation from a non-equilibrium gas-phase in exoplanetary atmospheres}

   \author{
          S. Kiefer \inst{1, 2, 3}
          \and
          H. Lecoq-Molinos \inst{2, 3, 1}
          \and
          Ch. Helling \inst{2, 3}
          \and
          N. Bangera \inst{2, 3}
          \and
          L. Decin \inst{1}
          }

   \institute{Institute of Astronomy, KU Leuven, Celestijnenlaan 200D, 3001 Leuven, Belgium\\
              \email{sven.kiefer@kuleuven.be}
         \and
             Space Research Institute, Austrian Academy of Sciences, Schmiedlstrasse 6, A-8042 Graz, Austria
         \and
             Institute for Theoretical Physics and Computational Physics, Graz University of Technology, Petersgasse 16 8010 Graz
             }

   \date{Received ...; accepted ...}

  \abstract
   {
   Recent observations suggest the presence of clouds in exoplanet atmospheres but have also shown that certain chemical species in the upper atmosphere might not be in chemical equilibrium. Present and future interpretation of data from, for example, CHEOPS, JWST, PLATO and Ariel requires a combined understanding of the gas-phase and the cloud chemistry.
   }
   {
   The goal of this work is to calculate the two main cloud formation processes, nucleation and bulk growth, consistently from a non-equilibrium gas-phase. The aim is further to explore the interaction between  a kinetic gas-phase and cloud micro-physics.
   }
   {
   The cloud formation is modelled using the moment method and kinetic nucleation which are coupled to a gas-phase kinetic rate network. Specifically, the formation of cloud condensation nuclei is derived from cluster rates that include the thermochemical data of (TiO$_2$)$_\mathrm{N}$ from $\mathrm{N}=1$ to $15$. The surface growth of 9 bulk Al/Fe/Mg/O/Si/S/Ti binding materials considers the respective gas-phase species through condensation and surface reactions as derived from kinetic disequilibrium. The effect of completeness of rate networks and the time evolution of the cloud particle formation is studied for an example exoplanet HD~209458~b.
   }
   {
   A consistent, fully time-dependent cloud formation model in chemical disequilibrium with respect to nucleation, bulk growth and the gas-phase is presented and first test cases are studied. This model shows that cloud formation in exoplanet atmospheres is a fast process. This confirms previous findings that the formation of cloud particles is a local process. Tests on selected locations within the atmosphere of the gas-giant HD~209458~b show that the cloud particle number density and volume reach constant values within 1s. The complex kinetic polymer nucleation of TiO$_2$ confirms results from classical nucleation models. The surface reactions of SiO[s] and SiO$_2$[s] can create a catalytic cycle that dissociates H$_2$ to 2 H, resulting in a reduction of the CH$_4$ number densities.
   }
   {}


   \maketitle
%

\section{Introduction}
\label{sec:Introduction}

    Clouds and dust are present in many astrophysical environments. In stellar environments, dust can be found in the outflows of asymptotic giant branch (AGB) stars \citep[e.g.][]{gail_dust_1984, fleischer_dynamical_1999, ferrarotti_composition_2006, hofner_dust_2009, gobrecht_dust_2016, khouri_study_2016, decin_study_2017} and Wolf–Rayet (WR) stars \citep{williams_infrared_1987, crowther_dust_2003, lau_nested_2022}. It can also be found in supernova \citep{tinyanont_supernova_2019, niculescu-duvaz_dust_2022, zhang_sn_2022}. Dust produced in these environments replenishes the interstellar medium (ISM) through radiation driven mass loss. In cooler objects, clouds are predicted within brown dwarfs \citep{allard_limiting_2001, ackerman_precipitating_2001, helling_dust_2004} and indirect evidence of clouds in brown dwarfs has been found \citep{maire_dusty_2020, ward-duong_gemini_2020, kammerer_gravity_2021}. Exoplanet atmospheres are also expected to have a strong cloud presence \citep{helling_exoplanet_2023}. Cloud particles typically have large opacities which lead to characteristically flat spectra in the optical and near infrared. And indeed, several exoplanet observations in these wavelength ranges show a flat spectrum \citep{bean_ground-based_2010, kreidberg_clouds_2014, espinoza_access_2019, spyratos_transmission_2021, libby-roberts_featureless_2022}. To understand cloud formation, one also needs to understand gas-phase chemistry and the nucleation process. In AGB stars, many chemical species have been detected \citep[e.g. AlF, MgNC, NaCN, CO, SiO, HCN, CS, PAHs;][]{highberger_heavy-metal_2001, decin_detection_2008, smolders_when_2010}. For exoplanets, recent medium and high resolution observations detected the presence of several atomic \citep[e.g. Mg, Na, Ca, Cr, Fe, Ni, V, Ti, Mn, O;][]{hoeijmakers_atomic_2018, hoeijmakers_hot_2020, prinoth_titanium_2022, borsa_high-resolution_2022} and molecular species \citep[e.g. CO, H$_2$O, CH4, NH$_3$, C$_2$H$_2$;][]{kok_detection_2013, hoeijmakers_medium-resolution_2018, guilluy_gaps_2022, guillot_giant_2022} pointing to a chemically rich environment.

    In collisionally dominated environments, the gas-phase can be modeled using chemical equilibrium models \citep{stock_fastchem_2018, woitke_ggchem_2021}. This is a good assumption for the deep atmosphere of exoplanets and brown dwarfs \citep{venot_better_2018} where the collisional timescales become small. In low density environments, like the outflow of AGB stars \citep{plane_master_2022, sande_role_2019} and the upper atmospheres of exoplanets \citep{rimmer_ionization_2013, baxter_evidence_2021, tsai_photochemically_2023, mendonca_three-dimensional_2018} and brown dwarfs \citep{helling_lightning_2019, lee_simplified_2020}, collisional timescales become large. In these environments, chemical disequilibrium processes, like radiation or quenching, can drive the gas-phase abundances out of equilibrium.

    The formation of clouds and dust starts with the formation of cloud condensation nuclei (CCNs). In gaseous exoplanets, CCNs cannot originate from the planet's surface, like in terrestrial planets, but have to be formed directly from the gas-phase through nucleation, which marks the transition from gas-phase chemistry to solid-phase chemistry. Classical nucleation theory (CNT) or modified classical nucleation theory (MCNT) is an often used approach to describe the rate at which CCNs are formed (nucleation rate). In order to describe the actual formation of clusters leading up to CCNs, a kinetic description can be used. \citep{patzer_dust_1998, lee_dust_2015, bromley_under_2016, boulangier_developing_2019, kohn_dust_2021, gobrecht_bottom-up_2022}. To calculate kinetic nucleation, the thermodynamic properties of clusters of the nucleating species have to be known. There are active efforts to derive the structures and properties of species that are associated with nucleation processes \citep{chang_inorganic_2005, chang_small_2013, patzer_density_2014, lee_dust_2015, gobrecht_bottom-up_2022, sindel_revisiting_2022, andersson_mechanisms_2023}. Even in the cases where data is available, it is often limited to the smallest cluster sizes because calculating the required properties becomes more computationally intensive for larger clusters \citep{sindel_revisiting_2022}.

    Supersaturated species can grow onto CCNs once they are present. There are two main ways these bulk growth processes can occur. Firstly there is condensation, which describes the deposition of gas-phase species onto CCNs (e.g. SiO $\rightarrow$ SiO[s]). Many models use condensation curves to determine where clouds can form \citep[e.g.][]{demory_inference_2013, webber_effect_2015, crossfield_observations_2015, kempton_observational_2017, roman_modeling_2017, roman_clouds_2021}. Secondly, bulk growth can occur through kinetic surface reactions \citep{patzer_dust_1998, helling_dust_2006, helling_modelling_2013}. In contrast to condensation, surface reactions include multiple chemical species to form the bulk material (e.g. SiS + H$_2$O → SiO[s] + H$_2$S ). In addition to providing additional bulk growth paths for condensing species, surface reactions also allow the bulk growth of materials which may not be stable in the gas-phase themselves. Both processes can be described kinetically \citep{patzer_dust_1998}.

    In this paper, we present a fully time-dependent description of the processes that lead to the formation of clouds in exoplanets, or dust in brown dwarfs, stars, and supernovae. We advance the nucleation description by modeling potential chemical pathways based on thermodynamic cluster properties and expand the kinetic description of surface reactions to chemical disequilibrium. With this model, we study the timescales of cloud formation in exoplanet atmospheres. The description of the chemical network, kinetic nucleation and bulk growth is given in Sect.~\ref{sec:Model}. The kinetic chemistry and kinetic nucleation are investigated in Sect.~\ref{sec:exploring}. The fully kinetic cloud formation model is then applied to temperature-pressure ($T_\mathrm{gas}$-$p_\mathrm{gas}$) points within the atmosphere of HD~209458~b (Sect.~\ref{sec:fully_kinetic}). Lastly, the summary is given in Sect.~\ref{sec:Conclusion}.

\section{Model}
\label{sec:Model}

    We present the models for the gas-phase, kinetic nucleation and bulk growth through condensation and surface reactions. In Sect.~\ref{sec:Model_chemistry}, we describe our chemical kinetics network of the gas-phase. The kinetic nucleation, which describes the time-dependent nucleation within disequilibrium environments, is described in Sect.~\ref{sec:Model_nucleation} (two-body reactions) and in Sect.~\ref{sec:app_threebody} (three-body reactions). In Sect.~\ref{sec:Model_condensation}, we describe the bulk growth through condensation and surface reactions. The derivation of the reaction supersaturation is given in Sect.~\ref{sec:Model_sr}. Finally, the connection between nucleation and bulk growth is described in Sect.~\ref{sec:Model_nuc_to_cond}.

    \subsection{Gas-phase chemistry}
    \label{sec:Model_chemistry}

    The evolution of the number density $n_i$ [cm$^{-3}$] of a given species $i$ is determined by the following equation:
    \begin{align}
        \frac{d n_i}{d t} =  \sum_{r \in F_i} \left( \nu_{i,r} k_r \prod_{j \in E_r} n_j \right) - \sum_{r \in D_i} \left( \nu_{i,r} k_r \prod_{j \in E_r} n_j \right) ,
    \end{align}
    where $F_i$ the set of reactions where the $i$-th species is a product, $D_i$ the set of reactions where the $i$-th species is a reactant, $E_r$ the set of reactants of reaction $r$, $n_j$ [cm$^{-3}$] the number densities of the reactants, $\nu_{i,r}$ the stoichiometric coefficient of the $i$-th species within reaction $r$, $k_r$ [cm$^{3(J_r-1)}$s$^{-1}$] the reaction rate for the reaction $r$ and $J_r$ the number of reactants in $E_r$. The sum over all $n_j$ is the total number density. The numerical solver is described in Appendix \ref{sec:Model_solver}.

    Chemical kinetic networks for the atmospheres of exoplanets include several hundred species and several thousand reactions \citep[e.g.][]{rimmer_chemical_2016, tsai_vulcan_2017, tsai_comparative_2021, venot_chemical_2012, venot_new_2020}. For this paper, we chose the "NCHO thermo network" of VULCAN\footnote{\url{https://github.com/exoclime/VULCAN/blob/master/thermo/NCHO_thermo_network.txt}}$^,$\footnote{A comparison to the chemical kinetics networks of \citet{decin_constraints_2018} and \citet{gobrecht_bottom-up_2022} can be found in Appendix \ref{sec:app_chemical_networks}.} \citep{tsai_vulcan_2017, tsai_comparative_2021}. This network includes 69 species and 780 reactions. In this work, TiO$_2$ is considered as the nucleation species. Because "NCHO thermo network" of VULCAN does not include reactions for the formation of TiO$_2$, we add several reactions from \citet{boulangier_developing_2019} leading to the formation of TiO$_2$. The selected gas-phase reactions can be found in Table \ref{tab:appendix_chemicalNetwork_nuccond}. Furthermore, we add reactions including Si species for the bulk growth species SiO and SiO$_2$ (see Sect.~\ref{sec:Model_condensation} and Table \ref{tab:appendix_chemicalNetwork_nuccond}).

     All calculations start from chemical equilibrium abundances calculated using \texttt{GGchem} \citep{woitke_ggchem_2021} for a solar-like composition \citep{asplund_chemical_2009}. A list of the considered species for the equilibrium calculation is given in Appendix \ref{sec:app_ggchem}.

    \subsection{Nucleation}
    \label{sec:Model_nucleation}
    Nucleation reaction networks are ideally constructed by considering multiple reaction pathways. Unfortunately, only few such studies exist \citep[see e.g.][]{bromley_under_2016, gobrecht_bottom-up_2022, andersson_mechanisms_2023}. In the kinetic network approach, the change in cluster number densities can be described as:
    \begin{align}
        \label{eq:Model_chemical_network}
        \frac{d n_\mathrm{N}}{d t} = \sum_{r \in F_\mathrm{N}} \left( \nu_{\mathrm{N},r} k_r^+ \prod_{j \in E_r} n_j \right) - \sum_{r \in D_\mathrm{N}} \left( \nu_{\mathrm{N},r} k_r^- \prod_{j \in E_r} n_j \right) ,
    \end{align}
    where $n_\mathrm{N}$ [cm$^{-3}$] is the number density of a given polymer of size N (also called N-mer), $F_\mathrm{N}$ is the set of forward reactions involving the N-mer, $D_\mathrm{N}$ is the set of backward reactions involving the N-mer. $k_r^+$ [cm$^{3(J_r-1)}$s$^{-1}$] is the forward reaction rate coefficient of reaction $r$, $k_r^-$ [cm$^{3(J_r-1)}$s$^{-1}$] is the backward reaction rate coefficient of reaction $r$ and $J_r$ is the number of reactants.

    We describe the growth reactions of nucleating species as two-body reactions (a + b $\rightarrow$ c). The forward reaction rate coefficient $k_r^+$ can then be described as \citep{peters_chapter_2017, boulangier_developing_2019}:
    \begin{align}
        \label{eq:Model_k}
        k_r^+ = \int_0^{\infty} \alpha_r(v_r) ~  \sigma_r(v_r) ~ v_r ~ f(v_r) ~ d v_r ,
    \end{align}
    where $v_r$ [cm s$^{-1}$] is the relative velocity of the collision partners, $\alpha_r (v_r)$ the sticking coefficient, $\sigma_r (v_r)$ [cm$^2$] is the reaction cross section and $f(v_r)$ the relative velocity distribution of the colliding particles. Similar to other work, we set the sticking coefficient $\alpha_j (\nu_r)$ to 1 because detailed values for nucleation reactions are not available so far \citep[e.g.][]{lazzati_non-local_2008, bromley_under_2016, boulangier_developing_2019}. The cross section is approximated by a collision of two hard spheres:
    \begin{align}
        \label{eq:Model_inelasitcCorssSec}
        \sigma_j = \pi (r_1 +r_2)^2 ,
    \end{align}
    where $r_1$ and $r_2$ [cm] are the interaction radii of the reaction partners. For this work, we consider radii including electrostatic forces \citep{kohn_dust_2021, gobrecht_bottom-up_2022, kiefer_effect_2023}. The relative velocity distribution is described by a Maxwell-Boltzmann distribution:
    \begin{align}
       \label{eq:Model_relMaxBoltz}
       f (\nu_r) = \left( \frac{\mu}{2 \pi k_B T_\mathrm{gas}} \right)^{3/2} 4 \pi v_r^2 \exp \left( \frac{\mu v_r^2}{2 k_B T}\right) ,
    \end{align}
    where $T_\mathrm{gas}$ [K] is the temperature, $k_B = 1.381 \times 10^{-23} \mathrm{erg}$ K$^{-1}$ is the Boltzmann constant and
    \begin{align}
        \label{eq:reduced_mass}
        \mu = \frac{m_1 m_2}{(m_1 + m_2)}
    \end{align}
    is the reduced mass [g] with m$_1$, m$_2$ [g] being the masses of the reaction partners. Solving the integral from Eq. \ref{eq:Model_k} yields:
   \begin{align}
        \label{eq:Model_k_integrated}
        k^+_j = \pi (r_1 + r_2)^2 \sqrt{\frac{8 k_B T_\mathrm{gas}}{\pi \mu}} .
   \end{align}
    The backwards reaction rate (c $\rightarrow$ a + b) is derived by assuming detailed balance. For this, an equilibrium state needs to be defined which we assume to be the chemical equilibrium state such that the law of mass action can be applied:
    \begin{align}
        \label{eq:Model_backwardsrate}
        k^-_j = k^+_j \frac{p^\ominus}{k_B T_\mathrm{gas}} \exp \left( \frac{G_{c}^\ominus(T_\mathrm{gas}) - G_{a}^\ominus(T_\mathrm{gas}) - G_{b}^\ominus(T_\mathrm{gas})}{k_B T_\mathrm{gas}} \right) ,
    \end{align}
    where $p^\ominus = 10^5$ Pa is the standard pressure and $G_{i}^\ominus(T_\mathrm{gas})$ [erg] is the Gibbs free energy of an $i$-mer at standard pressure.

    For this study, we consider TiO$_2$ as nucleating species. The chemical network is extended by including all forward (Eq. \ref{eq:Model_k}) and backward reactions (Eq. \ref{eq:Model_backwardsrate}) from the monomer up to the 15-mer. The Gibbs free energy data is taken from \citet{sindel_revisiting_2022}.

    \subsection{Three-body reactions for cluster formation}
    \label{sec:app_threebody}

    Three-body reactions are important for the formation of small clusters. On the one hand, third bodies can remove the energy of formation from an association (forward) reaction thus increasing cluster formation rates. On the other hand, collisions with third bodies can induce dissociation (backward) reactions. For this work, we consider three-body reactions for the cluster formation of TiO$_2$ up to (TiO$_2$)$_4$. The reaction rates are taken from \citet{kiefer_effect_2023} (see reactions number RNr 19 to 26 in Table \ref{tab:appendix_chemicalNetwork_nuccond}).

\begin{figure}
       \centering
       \includegraphics[width=\hsize]{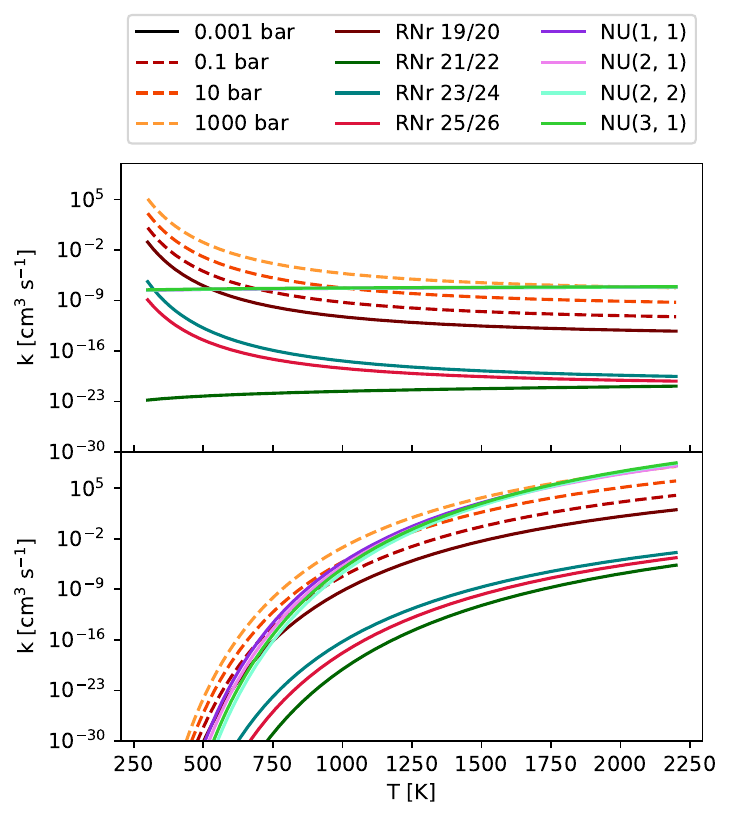}
       \caption{Reaction rate coefficients for the formation of TiO$_2$ clusters up to (TiO$_2$)$_4$. All solid lines assume $p_\mathrm{gas} = 0.002$ bar. \textbf{Top:} Association reactions. \textbf{Bottom:} Dissociation reactions.}
       \label{fig:app_3body}
    \end{figure}

    To analyse for which temperature and pressures three-body reactions dominate over two-body reactions, we compare the reaction rate coefficients. To allow a direct comparison, we multiply the reaction rate coefficients with the number density of third bodies. The comparison can be seen in Fig.~\ref{fig:app_3body}. For reaction RNr 19 and 20, multiple pressures are shown. Since all three-body reactions have the same pressure dependence, the scaling is the same for all of them. For two-body reactions, we use the following shorthand notation:
    \begin{align}
        \mathrm{NU}(i, j) &: (\mathrm{TiO_2})_i + (\mathrm{TiO_2})_i \leftrightarrow (\mathrm{TiO_2})_{(i+j)}
    \end{align}

    For the association and dissociation of (TiO$_2$)$_2$ from and into two monomers respectively (RNr 19/20 and NU(1, 1)), the three-body reaction becomes dominant below $T_\mathrm{gas} = 400$~K and for pressures higher than $p_\mathrm{gas} = 0.001$~bar. Above $T_\mathrm{gas} = 1300$~K and for pressures lower than $p_\mathrm{gas} = 10$~bar, the two-body reaction becomes dominant. In between, either type of reaction can be dominant. At higher pressures (e.g. $p_\mathrm{gas} > 1000$~bar), the three-body reaction becomes dominant even at high temperatures (e.g. $T_\mathrm{gas} > 2000$~K). For all other reactions compared in this section (RNr 21/22, 23/24, 25/26 and NU(2, 1), NU(2, 2), and NU(3, 1)), the two-body reaction typically dominates above $T_\mathrm{gas} > 400$~K. Three-body reactions only start to become important at very high pressures (e.g. $p_\mathrm{gas} > 1000$~bar).

    \subsection{Bulk growth}
    \label{sec:Model_condensation}

    In order to describe the bulk growth by gas-surface reaction we apply the moment method \citep{gail_primary_1986, gail_dust_1988, dominik_dust_1993, helling_dust_2001, helling_dust_2006}. The $j$-th moment $L_j$ [cm$^J$ g$^{-1}$] is defined as:
    \begin{align}
        \rho L_j = \int_{V_l}^\infty f(V) V^{j/3} dV,
    \end{align}
    where $j \in \{0, 1, 2, 3 \}$, $V$ [cm$^{3}$] is the cloud particle volume, $V_l$ [cm$^{3}$] is the minimum volume of a cloud particle to start bulk growth, $\rho$ [cm$^{-3}$] the gas density and $f(V)$ [cm$^{-6}$] is the cloud particle size distribution function. Using these moments the following cloud particle properties can be derived \citep{gail_dust_1988, helling_dust_2001}:
    \begin{align}
        n_d &= \rho L_0 ,\\
        \langle a \rangle &= \sqrt[3]{\frac{3}{4 \pi}} \frac{\rho L_1}{\rho L_0} ,\\
        \langle A \rangle &= \sqrt[3]{36 \pi} \frac{\rho L_2}{\rho L_0} ,\\
        \langle V \rangle &= \frac{\rho L_3}{\rho L_0} ,
    \end{align}
    where $n_d$ [cm$^{-3}$] is the cloud particle number density, $\langle a \rangle$ [cm] the mean cloud particle radius, $\langle A \rangle$ [cm$^2$] the mean cloud particle surface area and $\langle V \rangle$ [cm$^3$] the mean cloud particle volume. The change in the moments is determined by the nucleation and bulk growth and is described by the following set of equations\footnote{Gravitational settling and other transport processes may be added as source terms to the r.h.s., \citep{woitke_dust_2003}} \citep{helling_dust_2006}:
    \begin{align}
        \frac{\partial}{\partial t} \rho L_0 &= J_\star(V_l) ,\\
        \label{eq:L_not0_deq}
        \frac{\partial}{\partial t} \rho L_j &= V_l^{j/3} J_\star(V_l) + \frac{j}{3} \chi^\mathrm{net} \rho L_{j-1} ,\\
        \label{eq:chi_net}
         \chi^\mathrm{net} &= \sqrt[3]{36 \pi} \sum_{r \in C_d} {\Delta V_r n_r^\mathrm{key} \langle v_r \rangle \frac{\alpha_r (v_r)}{\nu_r^\mathrm{key}}} \left( 1- \frac{1}{(S_r)^{1/\nu_r^\mathrm{key}}} \frac{1}{b_r^\mathrm{surf}} \right) ,
    \end{align}
    where $J_\star(V_l)$ [cm$^{-3}$] is the nucleation rate (see Sect.~\ref{sec:Model_nuc_to_cond}), $C_d$ is the set of surface reactions, $\Delta V_r$ [cm$^{3}$] is the volume increase per surface reaction $r$, $n_r^\mathrm{key}$ [cm$^{-3}$] is the number density of the key gas-phase species, $\nu_r^\mathrm{key}$ is the stochiometirc coefficient of the key gas-phase species for the surface reaction $r$, $\langle v_r \rangle$ [cm s$^{-1}$] is the average relative velocity between the cloud particle and the key gas-phase species, $S_r$ is the reaction supersaturation (see section \ref{sec:Model_sr}) and $b_r^\mathrm{surf}$ the surface area fraction of the given bulk growth material. $\chi^\mathrm{net}$ [cm s$^{-1}$] is the net growth velocity. The key gas-phase species is the least abundant species involved in a given bulk growth reaction \citep{woitke_dust_2003, helling_dust_2006}. The left term in the bracket of Eq. \ref{eq:chi_net} represents the growth and the right term represents the evaporation. The cloud particle grows if the net sign of the bracket is positive and evaporates if the sign is negative. Similar to \citet{helling_dust_2006}, we assume that the surface area fraction can be approximated by the volume fraction:
    \begin{align}
        \label{eq:b_surf}
        b_r^\mathrm{surf} = \frac{A_r}{A_\mathrm{tot}} \approx \frac{V_r}{V_\mathrm{tot}} ,
    \end{align}
    where $A_\mathrm{tot}$ [cm$^2$] is the total cloud particle surface area and $V_\mathrm{tot}$ [cm$^3$] is the total cloud particle volume. The volume fraction of each cloud particle material is tracked separately.

    The growth of cloud particles through bulk growth depletes the gas-phase. Therefore, we adjust the number densities  $n_i$ of the species $i$ involved in the surface reaction $r$ \citep{helling_dust_2006}:
    \begin{align}
        \label{eq:cond_elementdepletion}
         \frac{d n_i}{dt} = \delta(i) ~\nu_{i,r} \frac{\chi^\mathrm{net}}{\Delta V_r} \rho L_2 ,
    \end{align}
    where $\delta(i)$ is equal to 1 for products and -1 for reactants.

    For this study, we consider TiO$_2$[s], Al$_2$O$_3$[s], SiO[s], SiO$_2$[s], MgO[s], Mg$_2$SiO$_4$[s], FeO[s], FeS[s], Fe$_2$O$_3$[s] and Fe$_2$SiO$_4$[s] as bulk growth materials and include the surface reactions listed in Table \ref{tab:con_reac}.

    \subsection{Reaction Supersaturation}
    \label{sec:Model_sr}
    To calculate the net growth velocity as described in Eq. \ref{eq:chi_net}, the reaction supersaturation needs to be calculated. Surface growth and evaporation of a material $s$ can occur via three types of reactions:
    \begin{align}
        &\text{\textbf{- Type 1: }} \mathrm{A(N-1)} + \mathrm{X} \leftrightarrows \mathrm{A(N)} \\
        &\text{\textbf{- Type 2: }} \mathrm{A(N-1)} + \mathrm{X} \leftrightarrows \mathrm{A(N)} + \mathrm{Y} \\
        &\text{\textbf{- Type 3: }} \mathrm{A(N-1)} + \sum_{X \in F} \nu_{X,r} ~\mathrm{X} \leftrightarrows \mathrm{A(N)} + \sum_{Y \in D} \nu_{Y,r} ~\mathrm{Y}
    \end{align}
    where X is a reactant, Y a product, $\mathrm{A(N)}$ a cloud particle containing N units of bulk growth material\footnote{$\mathrm{A(N)}$ can be heterogeneous and contain multiple bulk growth materials. To simplify notation, we only note the number of bulk growth units N of the bulk growth material involved in a given reaction.} (e.g. Mg$_2$SiO$_4$ would be 1 unit for Mg$_2$SiO$_4$[s]), $F$ the set of reactants, $D$ the set of products and $\nu_{i,r}$ the stochiometric coefficients of species $i$ for the surface reaction $r$. Type 1 reactions describe condensation, Type 2 reactions are chemical growth reactions, and Type 3 reactions involve surface chemistry. Cloud particles are assumed to be large enough for the following approximation to hold:
    \begin{align}
        \label{eq:big_appro}
        G^\ominus_\mathrm{A(N)} \approx G^\ominus_\mathrm{A(N-1)} + G^\ominus_\mathrm{A} ,
    \end{align}
    where $G^\ominus_\mathrm{A(N)}$ [erg] is the Gibbs free energy of formation at standard pressure of the $\mathrm{A(N)}$ cloud particle made from N units and $G^\ominus_\mathrm{A}$ [erg] is the Gibbs free energy of formation of a solid unit at standard pressure. The goal of this section is to find the reaction supersaturation $S_r$ of these reactions defined as:
    \begin{align}
        \label{eq:reac_s}
        S_r = \frac{R_f}{R_b} ,
    \end{align}
    where $R_f$ [s$^{-1}$] is the growth rate and $R_b$ [s$^{-1}$] is the evaporation rate. It is important to note that the following derivations are done in chemical disequilibrium. At no point in this section do we assume chemical equilibrium.

    \subsubsection{\texorpdfstring{$S_r$}{Lg} of Type 1 reactions (condensation)}

    Type 1 reactions are reactions where the reactant is also the bulk growth material. In this case the reaction supersaturation is equal to the supersaturation of the growth species:
    \begin{align}
        \label{eq:type1_sr}
        S_r = S = \frac{p_\mathrm{X}}{p_\mathrm{X}^\mathrm{vap}} = \frac{n_\mathrm{X}}{k_B T} \frac{1}{p^\ominus} \exp \left( \frac{G^\ominus_\mathrm{X} - G^\ominus_\mathrm{A}}{k_B T_\mathrm{gas}} \right) ,
    \end{align}
    where $p^\mathrm{vap}_\mathrm{X}$ [dyn cm$^{-2}$] is the vapour pressure of species X, $p_\mathrm{X}$ [dyn cm$^{-2}$] is the partial pressure of species X, $n_\mathrm{X}$ [cm$^{-3}$] the gas-phase number density of species X, $T_\mathrm{gas}$ [K] the temperature, and $G^\ominus_\mathrm{X}$ [erg] the Gibbs free energy of formation of the condensing species in the gas-phase at standard pressure. Since the bulk growth material exists in the gas-phase, the supersaturation is well defined.

    \subsubsection{\texorpdfstring{$S_r$}{Lg} of Type 2 reactions}

    Type 2 reactions were discussed in detail in previous studies \citep{gail_dust_1988, gauger_dust_1990, dominik_dust_1993, patzer_dust_1998}. The reaction supersaturation is given by \citep[adapted from][]{patzer_dust_1998}:
    \begin{align}
        S_r = \frac{n_\mathrm{X}}{n_\mathrm{Y}} \exp \left( \frac{G^\ominus_\mathrm{X} - G^\ominus_\mathrm{Y} - G^\ominus_\mathrm{A}}{k_B T_\mathrm{gas}} \right) ,
    \end{align}
    where $G^\ominus_\mathrm{X}$ [erg] and $G^\ominus_\mathrm{Y}$ [erg] are the Gibbs free energies of formation of the gas-phase reactant and product, respectively. $n_\mathrm{X}$ [cm$^{-3}$] is the number density of the reactant X and $n_\mathrm{Y}$ [cm$^{-3}$] is the number density of the product Y. Type 2 reactions are well defined even if the bulk growth material is not present in the gas-phase (see \citet{patzer_dust_1998} for further details).

    \subsubsection{\texorpdfstring{$S_r$}{Lg} of Type 3 reactions}

    Type 3 reactions involve surface chemistry. Considering these reactions is especially important if the bulk growth material is not present in the gas-phase. For example, Mg$_2$SiO$_4$[s] can condense via the surface reaction
    \begin{align}
        \label{eq:mg2sio4_reac}
        2\mathrm{Mg} + \mathrm{SiO} + 3\mathrm{H_2O} \leftrightarrow \mathrm{Mg_2SiO_4[s]} + 3\mathrm{H_2} .
    \end{align}
    To simplify the notation, we use the reaction of Eq. \ref{eq:mg2sio4_reac} as an example for this section and generalise the results in the end. For our example in this section, we assume Mg to be the key gas-phase species. The bulk growth rate of this reaction can then be described by \citep[adapted from][]{helling_dust_2006}:
    \begin{align}
        \label{eq:rf_mg2sio4}
        R_f = \left[ A_\mathrm{A(N-1)} v_\mathrm{key} \frac{1}{\nu_r^\mathrm{key}}\right] ~n_\mathrm{key} ,
    \end{align}
    where $A_\mathrm{A(N-1)}$ [cm$^2$] is the surface area of the $\mathrm{A(N-1)}$ cloud particle, $n_\mathrm{A(N-1)}$ [cm$^{-3}$] the number density of the $\mathrm{A(N-1)}$ cloud particle, $\nu_\mathrm{key}$ [cm s$^{-1}$] the relative velocity of the cloud particle and the key gas-phase species (e.g. Mg), and $n_\mathrm{key}$ [cm$^{-3}$] the number density of the key gas-phase species (e.g. Mg). Phase equilibrium for a given bulk growth reaction (short PGR, noted with $^\circ$) is characterised\footnote{In this section until and with Eq. \ref{eq:Sr_for_mg2sio4}, PGR is towards the reaction from Eq. \ref{eq:mg2sio4_reac} but the same derivation holds for an arbitrary reaction of type 3.} by the evaporation rate equalling the growth rate ($R_f^\circ = R_b^\circ$). Therefore according to Eq. \ref{eq:rf_mg2sio4}, the evaporation rate is:
    \begin{align}
        \label{eq:rb_mg2sio4}
        R_b &= \left[A_\mathrm{A(N-1)} v_\mathrm{key} \frac{1}{\nu_r^\mathrm{key}}\right] ~n_\mathrm{key}^\circ ,
    \end{align}
    where $n_\mathrm{key}^\circ$ [cm$^{-3}$] is the number density of the key gas-phase species in PGR (e.g. Mg). Since all non-key gas-phase species are typically much more abundant then the key gas-phase species, their number densities in non-PGR only slightly differ to their number densities in PGR. Therefore, the following approximation holds:
    \begin{align}
        \label{eq:one_reac_approx}
        \frac{n^\circ_\mathrm{SiO}}{n_\mathrm{SiO}} \approx \frac{n^\circ_\mathrm{H_2O}}{n_\mathrm{H_2O}} \approx \frac{n^\circ_\mathrm{H_2}}{n_\mathrm{H_2}} \approx 1 .
    \end{align}
    This allows us to write the reaction supersaturation as:
    \begin{align}
        S_r^2 &= \frac{(n_\mathrm{Mg})^2}{(n^\circ_\mathrm{Mg})^2} = \frac{(n_\mathrm{Mg})^2}{(n^\circ_\mathrm{Mg})^2} \frac{n^\circ_\mathrm{SiO} (n^\circ_\mathrm{H_2O})^3 (n_\mathrm{H_2})^3}{n_\mathrm{SiO} (n_\mathrm{H_2O})^3 (n^\circ_\mathrm{H_2})^3} \left[ \frac{n_\mathrm{SiO} (n_\mathrm{H_2O})^3 (n^\circ_\mathrm{H_2})^3} {n^\circ_\mathrm{SiO} (n^\circ_\mathrm{H_2O})^3 (n_\mathrm{H_2})^3} \right] \\
        &\approx \frac{(n_\mathrm{Mg})^2 n_\mathrm{SiO} (n_\mathrm{H_2O})^3}{(n_\mathrm{H_2})^3} \frac{(n^\circ_\mathrm{H_2})^3}{(n^\circ_\mathrm{Mg})^2 n^\circ_\mathrm{SiO} (n^\circ_\mathrm{H_2O})^3} ,
    \end{align}
    leaving only the ratios of gas number densities in PGR to find.

    We start the derivation with a thought experiment. Imagine a box containing a given elemental abundance of Mg, SiO, H$_2$O, H$_2$, $\mathrm{A(N-1)}$ and $\mathrm{A(N)}$. We assume that in this box, gas-phase species only react with each other via the specific surface reaction from Eq. \ref{eq:mg2sio4_reac} but do not react with each other otherwise. Over time, the box will evolve towards PGR for this specific chemical configuration. In PGR, the entropy of the box will be maximised which is equivalent to minimising the Gibbs free energy for the reactants and products of the given reaction. This minimisation problem with its constraints can be expressed in the following Lagrangian function:
    \begin{align}
        \nonumber\mathcal{L} = &\left[ \sum_{j \in E} N_{j} G^\ominus_{j} + N_{j} k_B T_\mathrm{gas} \ln \left( \frac{N_j}{N}\right) \right] \\
        \nonumber &+ \lambda_1(C_1 +   N_\mathrm{Mg} - 2 N_\mathrm{SiO}) \\
        \nonumber &+ \lambda_2(C_2 + 3 N_\mathrm{Mg} - 2 N_\mathrm{H_2O})\\
        \nonumber &+ \lambda_3(C_3 +   N_\mathrm{Mg} - 2 N_\mathrm{A(N-1)})\\
        \nonumber &+ \lambda_4(C_4 - 3 N_\mathrm{Mg} - 2 N_\mathrm{H_2})\\
                  &+ \lambda_5(C_5 -   N_\mathrm{Mg} - 2 N_\mathrm{A(N)}) ,
    \end{align}
    where $E=\{\mathrm{Mg}, \mathrm{SiO}, \mathrm{H_2O}, \mathrm{H_2}, \mathrm{A(N-1)}, \mathrm{A(N-1)} \}$ is the set of particles, $N_j$ the total number of particles $j$, $N$ is the total number of gas particles, $G^\ominus_j$ [erg] the Gibbs free energy of formation of particle $j$ at standard pressure, $\lambda_i$ [erg] are the Lagrangian multipliers and $C_i$ are constants. The constrains from $\lambda_1$, $\lambda_2$, and $\lambda_3$ are keeping the ratio of Mg, SiO, H$_2$O and $\mathrm{A(N-1)}$ per reaction constant using Mg as reference. The constraints from $\lambda_4$, and $\lambda_5$ ensure mass conservation. Minimizing this Lagrangian for all molecules and cloud particles results in the following set of equations:
    \begin{align}
        \frac{d \mathcal{L}}{d N_\mathrm{Mg}} = &G^\ominus_\mathrm{Mg} + k_B T_\mathrm{gas} \ln \left( \frac{N_\mathrm{Mg}}{N}\right) + \lambda_1 + 3 \lambda_2 + \lambda_3 - 3 \lambda_4 - \lambda_5 ,\\
        \frac{d \mathcal{L}}{d N_\mathrm{SiO}} = &G^\ominus_\mathrm{SiO} + k_B T_\mathrm{gas} \ln \left( \frac{N_\mathrm{SiO}}{N}\right) - 2 \lambda_1 ,\\
        \frac{d \mathcal{L}}{d N_\mathrm{H_2O}} = &G^\ominus_\mathrm{H_2O} + k_B T_\mathrm{gas} \ln \left( \frac{N_\mathrm{H_2O}}{N}\right) - 2 \lambda_2 ,\\
        \frac{d \mathcal{L}}{d N_\mathrm{A(N-1)}} = &G^\ominus_\mathrm{A(N-1)} + k_B T_\mathrm{gas} \ln \left( \frac{N_\mathrm{A(N-1)}}{N}\right) - 2 \lambda_3 ,\\
        \frac{d \mathcal{L}}{d N_\mathrm{H_2}} = &G^\ominus_\mathrm{H_2} + k_B T_\mathrm{gas} \ln \left( \frac{N_\mathrm{H_2}}{N}\right) - 2 \lambda_4 ,\\
        \frac{d \mathcal{L}}{d N_\mathrm{A(N)}} = &G^\ominus_\mathrm{A(N)} + k_B T_\mathrm{gas} \ln \left( \frac{N_\mathrm{A(N)}}{N}\right) - 2 \lambda_5 .
    \end{align}
    In PGR, the Lagrangian function is minimal and thus the derivatives are zero. Solving this set of equations, using the approximation of Eq. \ref{eq:big_appro}, and going from particle numbers $N_j$ to particle number densities $n_j$ leads to:
    \begin{align}
        \label{eq:gibbs_for_cond}
        \nonumber\frac{(n^\circ_\mathrm{Mg})^2 n^\circ_\mathrm{SiO} (n^\circ_\mathrm{H_2O})^3}{(n^\circ_\mathrm{H_2})^3} \approx &\left(\frac{p^\ominus}{k_B T_\mathrm{gas}}\right)^{3} \exp \left( \frac{-1}{k_B T_\mathrm{gas}} \Big[ \right.2 G_\mathrm{Mg}^\ominus + G_\mathrm{SiO}^\ominus \\
        &\left.\left. \quad   + 3 G_\mathrm{H_2O}^\ominus - 3 G_\mathrm{H_2}^\ominus - G_\mathrm{Mg_2SiO_4[S]}^\ominus \right] \right) .
    \end{align}
    This result gives us the relation between the number densities of the reactants and products of the bulk growth reaction. Hence, the reaction supersaturation for the reaction of Eq. \ref{eq:mg2sio4_reac} is given by:
    \begin{align}
        \label{eq:Sr_for_mg2sio4}
        \nonumber S_r^2 &\approx \frac{(n_\mathrm{Mg})^2 n_\mathrm{SiO} (n_\mathrm{H_2O})^3}{(n_\mathrm{H_2})^3} \left(\frac{p^\ominus}{k_B T_\mathrm{gas}}\right)^{-3} \\
         &\quad \exp \left( \frac{1}{k_B T_\mathrm{gas}} \left[2 G_\mathrm{Mg}^\ominus + G_\mathrm{SiO}^\ominus + 3 G_\mathrm{H_2O}^\ominus -  3 G_\mathrm{H_2}^\ominus - G_\mathrm{Mg_2SiO_4[S]}^\ominus \right] \right) .
    \end{align}
    For an arbitrary type 3 reaction that is limited by a key gas-phase species, the reaction supersaturation is then given by:
    \begin{align}
        \nonumber S_r^{\nu_\mathrm{key}} = &\frac{\Pi_{X \in F} n_X^{\nu_X}}{\Pi_{Y \in D} n_Y^{\nu_Y}} \left(\frac{p^\ominus}{k_B T_\mathrm{gas}}\right)^{l_Y-l_X} \\
        &\exp \left( \frac{1}{k_B T_\mathrm{gas}} \left[G_\mathrm{A}^\ominus - \sum_{X \in F} \nu_X G_\mathrm{X}^\ominus + \sum_{Y \in D} \nu_Y G_\mathrm{Y}^\ominus \right] \right) ,
    \end{align}
    where $l_X$ is the number of reactants and $l_Y$ is the number of products. If only 1 reactant is considered, this result matches the result for type 1 reactions. It also matches type 2 reactions if only 1 key reactant and 1 key product are considered. In the case of chemical equilibrium, our description of the supersaturation ratio for type 3 reactions is the same as the one found by \citet{helling_dust_2006}.

    We define the right hand side of Eq. \ref{eq:gibbs_for_cond} as the reaction vapor coefficient $c_r^\mathrm{vap}$ [cm$^{-3(l_x - l_y)}$] and fit it with:
    \begin{align}
        \label{eq:s_fit}
        \ln \left( \frac{c_r^\mathrm{vap}}{[\mathrm{cm}^{-3(l_x - l_y)}]} \right) = \sum_{i=0}^3 \frac{s_i}{T_\mathrm{gas}^i} .
    \end{align}
    The fitting parameters for the surface reactions considered in this paper are given in Table \ref{tab:con_reac}. This allows us to write the reaction supersaturation as:
    \begin{align}
        S_r^{\nu_\mathrm{key}} &= \frac{\Pi_{X \in F} n_X^{\nu_X}}{\Pi_{Y \in D} n_Y^{\nu_Y}} \frac{1}{c_r^\mathrm{vap}} .
    \end{align}

    \subsection{Formation rate of CCNs}
    \label{sec:Model_nuc_to_cond}

    To describe nucleation kinetically, we require the properties of each considered cluster size up until the size N$_\star$ where the clusters become preferably thermally stable. For this, the Gibbs free energies $G_{\mathrm{N}}$ (Eq. \ref{eq:Model_backwardsrate}), the interaction radii $r_\mathrm{N}$ (Eq. \ref{eq:Model_inelasitcCorssSec} and \ref{eq:Model_k_integrated}) and the masses $m_\mathrm{N}$ (Eq. \ref{eq:reduced_mass}) of all considered cluster sizes need to be known. Unfortunately, only limited data is available \citep[see for example the case of TiO$_2$:][]{qu_theoretical_2006, lundqvist_dft_2006, zhai_probing_2007, chiodo_structure_2011, lee_dust_2015, sindel_revisiting_2022}. We use the data of \citet{sindel_revisiting_2022} and approximate the nucleation rate with the largest cluster for which all necessary data is available, which in this case is N$_\mathrm{max} = 15$. The quality of this assumption for N$_\mathrm{max}$ is tested in Sect.~\ref{sec:res_crit}.

    We follow the approach of \citet{patzer_dust_1998} and calculate the nucleation rate as:

    \begin{align}
        \label{eq:nuc_rate}
        J_\star(V_l) = \frac{d n_{\mathrm{N}_\mathrm{max}}}{dt} ,
    \end{align}
    which describes the rate at which CCNs are formed. The growth from clusters made from N$_\mathrm{max}$ monomers to clusters with a volume of $V_l$ depletes the gas-phase of the clustering species. Because polymer nucleation is considered, all N-mers up to a given $Q$-mer are depleted:
    \begin{align}
        \frac{d n_\mathrm{N}}{dt} &= - J_\star(V_l)  \frac{n_\mathrm{N}}{\omega_m} , \\
        \omega_m &= \frac{V_l}{\Delta V_r} \sum_{\mathrm{N}=1}^Q \mathrm{N} ~n_\mathrm{N} ,
    \end{align}
    where $n_\mathrm{N}$ [cm$^{-3}$] are N-mer number densities and $1 \leq \mathrm{N} \leq Q$. For TiO$_2$, the change in the number density of clusters of size N$_\mathrm{max} = 15$ defines nucleation (see Eq. \ref{eq:nuc_rate}). We therefore exclude this size in the depletion description and select $Q_\mathrm{TiO_2} = 14$.

    Starting from $V_l$, the particles can grow via surface growth. Similarly, they can shrink via evaporation down to size $V_l$. To numerically separate between evaporation and nucleation, cloud particles should only evaporate down to size $V_l$. For the ODE solver, we need a continuous transition and therefore adjust the evaporating surface area by multiplying the evaporation term of Eq. \ref{eq:chi_net} with
    \begin{align}
        \label{eq:xi}
        c_\mathrm{vol} = \left( \frac{\rho L_3}{\rho L_0} - V_l \right) \frac{\rho L_0}{\rho L_3} .
    \end{align}
    Including Eq. \ref{eq:xi} in Eq. \ref{eq:chi_net} results in:
    \begin{align}
        \label{eq:real_chi}
        \chi^\mathrm{net} &= \sqrt[3]{36 \pi} \sum_{r \in C_d} {\Delta V_r n_r^\mathrm{key} \langle v_r \rangle} \frac{\alpha_r (v_r)}{\nu_r^\mathrm{key}} \left( 1- \frac{c_\mathrm{vol}}{S_r^{1/\nu_r^\mathrm{key}} b_r^\mathrm{surf}} \right) .
    \end{align}
    For the rest of this paper, we are using Eq. \ref{eq:real_chi} for $\chi^\mathrm{net}$ in Eqs. \ref{eq:L_not0_deq} and \ref{eq:cond_elementdepletion}.

\section{Exploring kinetic chemistry and nucleation}
\label{sec:exploring}

    In Sect.~\ref{sec:res_chem}, we create a chemical network for the kinetic cloud formation model. Using this network, we study the impact on the gas phase when combining different chemical kinetics networks, nucleation and bulk growth. Describing bulk growth fully time dependent revealed a SiO-SiO$_2$ cycle within the surface reactions which is discussed in Sect.~\ref{sec:res_ch4}. In Sect.~\ref{sec:res_crit}, we evaluate the nucleation rate's dependence on different maximum cluster sizes N$_\mathrm{max}$.

    \subsection{A chemical network for kinetic cloud formation}
    \label{sec:res_chem}

\begin{figure}[!h]
       \centering
       \includegraphics[width=\hsize]{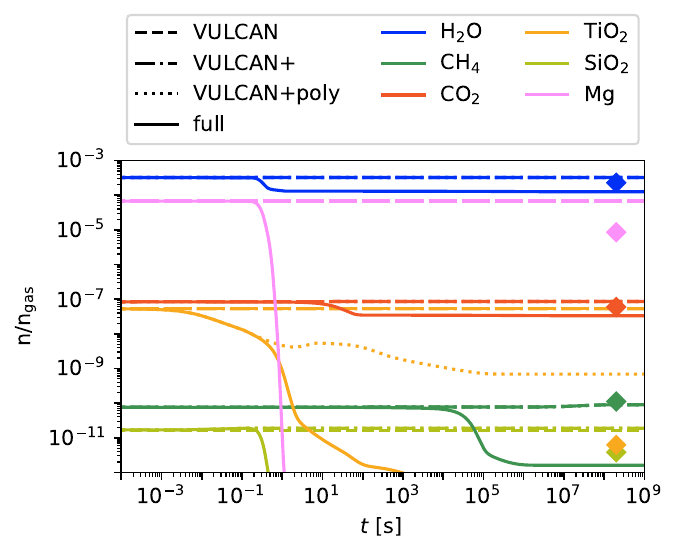}
       \caption{Concentrations of selected gas-phase species for $p_\mathrm{gas}= 0.002$~bar at $T_\mathrm{gas}= 1378$~K using different chemical kinetics networks. The diamond shaped marker show \texttt{GGchem} results including equilibrium condensation.}
       \label{fig:network_study}
    \end{figure}

    To find the impact of combining different chemical networks, nucleation and bulk growth on the number densities of the gas-phase species, we compare the gas-phase abundances in 4 cases:
    \begin{itemize}
        \item \textbf{Equilibrium:} Chemical equilibrium number densities calculated using \texttt{GGchem}. This calculation is time independent. \\

        \item \textbf{VULCAN:} The "NCHO thermo network" reactions of VULCAN. \\

        \item \textbf{VULCAN+:} This network combines the "NCHO thermo network" reactions of VULCAN, and the Ti and Si reactions as listed in Table \ref{tab:appendix_chemicalNetwork_nuccond}\\

        \item \textbf{VULCAN+poly:} This network combines the "NCHO thermo network" reactions of VULCAN, the Ti and Si reactions as listed in Table \ref{tab:appendix_chemicalNetwork_nuccond}, the polymer nucleation reactions for TiO$_2$ as described in Sect.~\ref{sec:Model_nucleation}, and the formation of TiO$_2$ CCNs as described in Sect.~\ref{sec:Model_nuc_to_cond}. \\

        \item \textbf{Full:} This network combines the "NCHO thermo network" reactions of VULCAN, the Ti and Si reactions as listed in Table \ref{tab:appendix_chemicalNetwork_nuccond}, the polymer nucleation reactions for TiO$_2$ as described in Sect.~\ref{sec:Model_nucleation}, the formation of TiO$_2$ CCNs as described in Sect.~\ref{sec:Model_nuc_to_cond}, and the bulk growth through condensation and surface reactions as described in Sect.~\ref{sec:Model_condensation}.
    \end{itemize}
    All simulations start from equilibrium number densities calculated with \texttt{GGchem}. All time axis in this paper measure the evaluation time $t$ [s] of the disequilibrium chemistry starting from equilibrium conditions at $t =0$ s. The results for the $T_\mathrm{gas}$-$p_\mathrm{gas}$ point\footnote{This $T_\mathrm{gas}$-$p_\mathrm{gas}$ point corresponds to a location within the evening terminator of HD~209458~b.} $p_\mathrm{gas}= 0.002$ bar at $T_\mathrm{gas}= 1378$ K can be seen in Fig. \ref{fig:network_study}. The chemical equilibrium number densities match the predicted number densities of VULCAN and are therefore not shown in the figure.

    \begin{table*}
        \centering
        \caption{Offset values between the number densities of given molecular species for different chemical networks according to Eq. \ref{eq:chem_compare}.}
        \label{tab:res_combining_chemical}
        \begin{tabular}{l l l l l l l l l}
            \hline\hline
                                                            & H$_2$    & H$_2$O   & CO$_2$   & CH$_4$   & TiO$_2$  & SiO$_2$  & Mg   \\
             \hline \hline
             $P(A, \mathrm{Equilibrium}, \mathrm{VULCAN})$  & 4.84e-05 & 7.35e-03 & 1.58e-02 & 6.35e-02 & -        & -        & -    \\
             $P(A, \mathrm{VULCAN}, \mathrm{VULCAN+})$      & 5.18e-11 & 1.34e-07 & 1.83e-05 & 1.18e-07 & 7.43e-04 & 5.25e-02 & -    \\
             $P(A, \mathrm{VULCAN+}, \text{VULCAN+poly})$   & 4.99e-08 & 1.32e-04 & 1.49e-04 & 1.32e-04 & 1.89     & 1.32e-04 & -    \\
             $P(A, \text{VULCAN+poly}, \mathrm{full})$      & 1.49e-03 & 0.41     & 0.41     & 1.74     & 5.21     & 2.35     & 12.8 \\
             \hline
        \end{tabular}
    \end{table*}

    To compare the difference in the predicted number densities, we compare the maximum absolute difference between the logarithm of the number densities of species $A$ between two chemical networks $C_1$ and $C_2$:
    \begin{align}
        \label{eq:chem_compare}
        P(A, C_1, C_2) = \max \{ |\log_{10}(n_{C_1}^{A}(t) / n_{C_2}^{A}(t))|, \forall t \in [ 10^{-4}, 10^{9} ]  \}.
    \end{align}
    The $P$ values for $p_\mathrm{gas}= 0.002$ bar and $T_\mathrm{gas}= 1378$ K, which represents a low-pressure level in a relatively cool atmosphere, are shown in Table \ref{tab:res_combining_chemical}. For the comparison we chose H$_2$ because it is the dominant gas phase species, H$_2$O and CO$_2$ because they are commonly studied, methane (CH$_4$) because it has distinct spectral features, Mg and SiO$_2$ because they are condensing species, and TiO$_2$ because it is the nucleating species.

    Comparing Equilibrium to VULCAN and VULCAN to VULCAN+ shows in both cases close to no difference ($P < 0.1$) in the predicted number densities of H$_2$, H$_2$O, CO$_2$, CH$_4$, and SiO$_2$. In VULCAN+poly TiO$_2$ nucleation reactions and the formation of TiO$_2$ CCNs are added. Therefore, it comes at no surprise that the number density of TiO$_2$ decreases by close to two orders of magnitude. The impact for H$_2$, H$_2$O, CO$_2$, CH$_4$, and SiO$_2$ on the other hand is still negligible. When bulk growth is added in the full network, the number densities of TiO$_2$, SiO$_2$ and Mg decrease by several orders of magnitude due to being depleted by the bulk growth processes. The impact of bulk growth can also be seen in the number densities of H$_2$O, CO$_2$, and CH$_4$. The change in CH$_4$ is discussed in Sect.~\ref{sec:res_ch4}. The number density of H$_2$ on the other hand is not significantly affected.

    Also shown in Fig.~\ref{fig:network_study} are the gas-phase concentrations calculated using \texttt{GGchem} including the equilibrium condensation of the bulk grow materials. Compared to our results, \texttt{GGchem} equilibrium condensation results predict higher gas-phase concentrations in the cloud forming species TiO$_2$, SiO$_2$, and Mg as well as the gas-phase only species H$_2$O, CH$_4$, and CO$_2$. Because our work treats cloud formation kinetically, these differences can be caused by the nucleation or surface reactions which are both not considered within \texttt{GGchem}. In environments like the ISM or Titan's atmosphere, surface reactions are known to cause deviations from equilibrium gas-phase abundances (see Sect.~\ref{sec:res_ch4}).

    Selecting a suitable gas-phase chemical kinetics network is important. We chose VULCAN because it includes commonly considered species such as H$_2$O, CO$_2$ and CH$_4$ with a reasonably low number of reactions (780). Other chemical kinetics networks for exoplanet atmospheres include thousands of reactions \citep[e.g.][]{venot_chemical_2012, venot_chemical_2015, rimmer_chemical_2016, hobbs_chemical_2019, venot_new_2020}. Because of the computational intensity of these networks, their evaluation is often limited to 1D models \citep[e.g.][]{moses_photochemistry_2005, chadney_effect_2017, hobbs_molecular_2022, barth_moves_2021}. Adding kinetic nucleation and bulk growth to the chemical network can increase the computational time considerably. For the simulations in this section, the evaluation time doubled if nucleation and bulk growth were considered. If enough computational resources are available, our kinetic nucleation and bulk growth model can be combined with extensive chemical networks for a detailed study. Furthermore, models and observations have shown that the 3D structure of exoplanets can have an impact on the gas-phase chemistry \citep{baeyens_grid_2021, prinoth_titanium_2022, lee_mini-chemical_2023}. To evaluate the chemistry and cloud formation within 3D models, the cloud formation description can be combined with small but accurate networks \citep{tsai_mini-chemical_2022}.

    Most chemical kinetics networks for exoplanet atmospheres do not include many Mg, Ti, Si, or Fe bearing species. In our simulations, only the surface reactions c0, c1, c2, c4,  c16, c71, c80, c85, c101, and c104 (see Table \ref{tab:con_reac}) have all reactants and products within the chemical kinetics network. All other bulk growth reactions relay at least partially on gas-phase species which are only calculated in equilibrium. Ideally, all reactants and products of surface reactions should be included in the chemical kinetics network but they are not always available in literature and would drastically increase the number of reactions.

    \subsection{\texorpdfstring{SiO-SiO$_2$}{Lg} cycle}
    \label{sec:res_ch4}
    In a fully kinetic description of cloud formation, minor species are affected as well. As shown in Fig. \ref{fig:network_study}, adding bulk growth reactions significantly impacts the number density of CH$_4$, even though it is not directly involved in any bulk growth processes. This is important since CH$_4$ is discussed as bio molecule in terrestrial atmosphers \citep{thompson_case_2022, huang_methanolpoor_2022} and is observable in hot Jupiter atmospheres \citep{swain_water_2009, barman_simultaneous_2015, guilluy_exoplanet_2019}.

    The 4 bulk growth reaction causing the decrease in CH$_4$ are c17, c19, c42, and c44. A closer look at these surface reactions revealed the following cycle:
    \begin{align}
       \mathrm{c42:}&& 2~\mathrm{SiO} &\rightarrow 2 \mathrm{SiO[s]}, \\
       \mathrm{c44:}&& 3~\mathrm{H_2} + 2~\mathrm{SiO[s]} &\rightarrow 2~\mathrm{SiH} + 2~\mathrm{H_2O}, \\
       \mathrm{c19:}&& 4~\mathrm{H_2O} + 2~\mathrm{SiH} &\rightarrow 2~\mathrm{SiO_2[s]} + 4~\mathrm{H_2} + 2~\mathrm{H}, \\
       \mathrm{c17:}&& 2~\mathrm{H_2} + 2~\mathrm{SiO_2[s]}  &\rightarrow 2~\mathrm{SiO} + 2~\mathrm{H_2O} .
    \end{align}
    The net process of this cycle is:
    \begin{align}
        \mathrm{H_2} \rightarrow 2~\mathrm{H}.
    \end{align}
    This additional pathway for the dissociation of H$_2$ to H decreases the number density of CH$_4$ through the following reactions:
    \begin{align}
        \mathrm{CH_4} + \mathrm{H} &\rightarrow \mathrm{CH_3} + \mathrm{H_2}, \\
        \mathrm{CH_3} + \mathrm{H} &\rightarrow \mathrm{CH_2} + \mathrm{H_2}, \\
        \mathrm{CH_2} + \mathrm{H} &\rightarrow \mathrm{CH} + \mathrm{H_2} .
    \end{align}
   Most of the carbon from CH$_4$ is transferred into H$_2$CO with the following reaction:
    \begin{align}
        \mathrm{CH} + \mathrm{H_2O} &\rightarrow \mathrm{H_2CO} + \mathrm{H}.
    \end{align}
    The change in carbon chemistry then also impacts other species like HCN, HCO, and C$_2$H$_2$.

\begin{figure}
       \centering
       \includegraphics[width=\hsize]{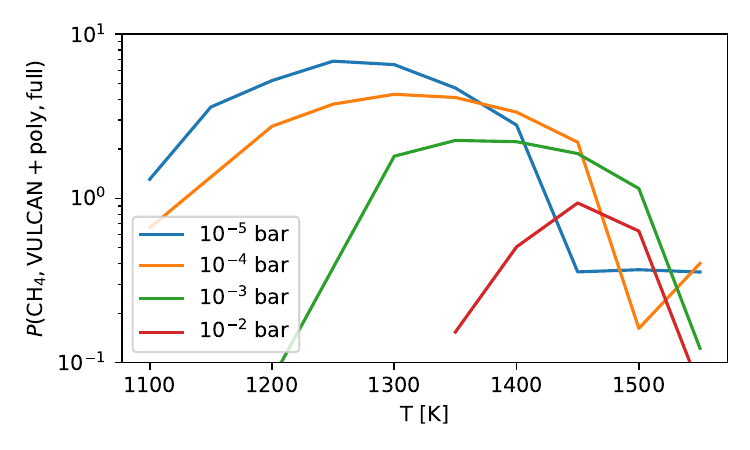}
       \caption{Differences in the CH$_4$ abundance between chemical equilibrium and the VULCAN+poly network for various temperatures and pressures.}
       \label{fig:res_ch4inv}
    \end{figure}

    To investigate at which pressures and temperatures the SiO-SiO$_2$ cycle significantly impacts the CH$_4$ abundance, $P(\mathrm{CH}_4, \mathrm{Equilibrium}, \mathrm{VULCAN+poly})$ values for a range of pressures and temperatures were calculated. The results are shown in Fig.~\ref{fig:res_ch4inv}. The difference in the CH$_4$ abundance is largest for pressures lower than $p_\mathrm{gas} < 10^{-3}$ bar and for temperatures around 1300~K.

    The surface reactions used in this work were derived using a stoichiometric argument \citep{helling_dust_2006, helling_dust_2008, helling_dust_2017, helling_sparkling_2019}. Unfortunately, more detailed studies of surface reactions of bulk growth materials are missing in literature. To determine if all surface reactions of the SiO-SiO$_2$ cycle are likely to occur in exoplanet atmospheres, further investigations are needed. Because other processes, like quenching or photo-chemistry, can have similar effects \citep{moses_disequilibrium_2011}, it will be difficult to gain insights into surface reactions through observations.

    \citet{molaverdikhani_role_2020} analysed the impact of clouds on the methane abundance. They found that clouds can increase the temperature in the photo-sphere which in turn reduces the methane abundance. In contrast to our work, they used condensation curves rather than surface reactions and therefore they did not observe a direct impact of cloud formation on the CH$_4$ abundance.

    In the (ISM) surface reaction are discussed as sources for molecular hydrogen in the gas phase \citep{hollenbach_surface_1971, williams_gas_2005, sabri_interslar_2013, dishoeck_astrochemistry_2014, herbst_three_2014, herbst_synthesis_2017}. Similar to our study, in the ISM the surface of dust grains act as a catalyst but in contrast to our work, they do not result in bulk material being deposited. The calculation of ISM surface reaction rates typically accounts for the vibrational frequency of the reactants and the energy barriers between different sites on the dust particle \citep{dishoeck_astrochemistry_2014}. Our surface reaction description could be improved by similar considerations. However, the large number of surface reactions considered and the heterogeneity of the cloud particle make such evaluations difficult.

    Similar to the ISM, the surfaces of aerosols in Titan's atmosphere can enhance the recombination of H into H$_2$ \citep{courtin_uv_1991, bakes_role_2003, sekine_role_2008}. In addition, gas-phase catalytic cycles using hydrocarbons for the hydrogen recombination are postulated \citep{yung_photochemistry_1984, toublanc_photochemical_1995, lebonnois_atomic_2003}. Both effects change the atomic hydrogen abundance which, similar to our work, can affect the abundance of hydrocarbons in return. In contrast to our work, the catalytic cycles considered are gas-phase only and do not result in bulk material being deposited.

    \subsection{Kinetic nucleation of \texorpdfstring{TiO$_2$}{Lg}}
    \label{sec:res_crit}

    \begin{figure}
       \centering
       \includegraphics[width=\hsize]{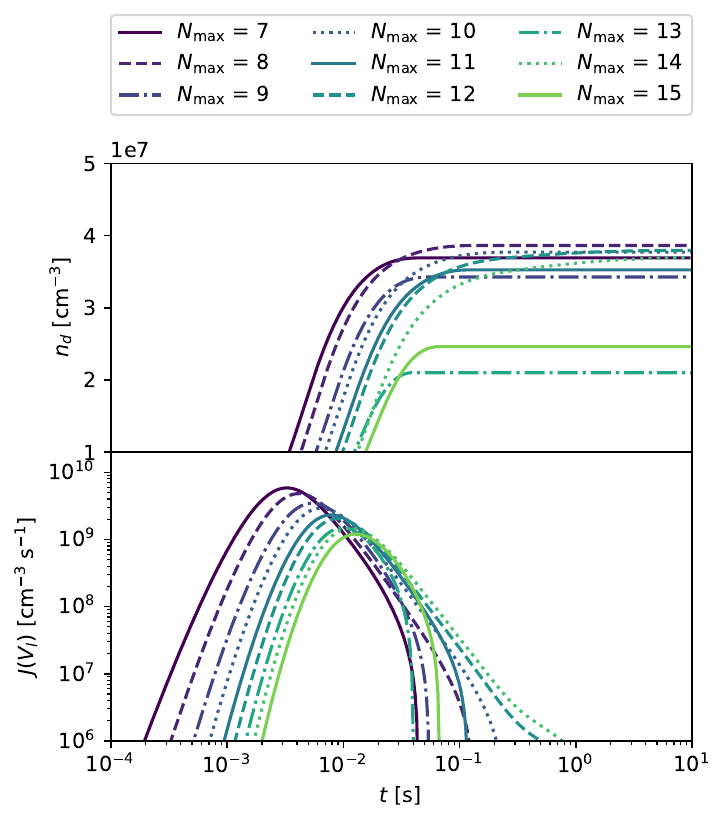}
       \caption{Cloud particle number densities (\textbf{Top}) and nucleation rates (\textbf{Bottom}) for TiO$_2$ nucleation with different N$_\mathrm{max}$ at $p_\mathrm{gas}= 0.02$ bar and $T_\mathrm{gas}= 1379$ K.}
       \label{fig:res_nucpersize}
    \end{figure}

    \begin{figure}
       \centering
       \includegraphics[width=\hsize]{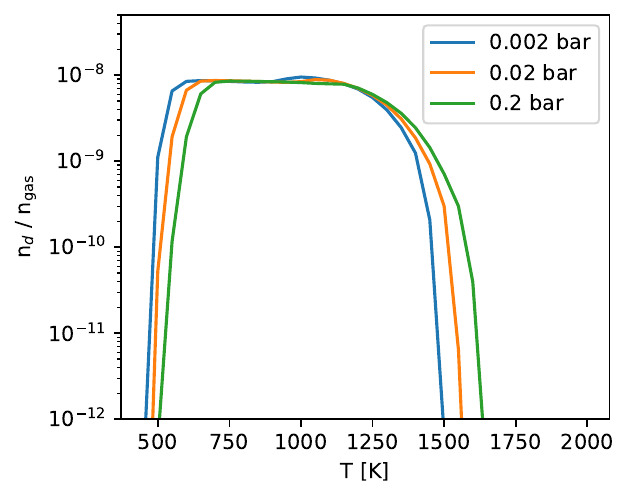}
       \caption{Cloud particle concentrations for TiO$_2$ nucleation for a range of temperatures and pressures.}
       \label{fig:res_nucpertemp}
    \end{figure}

    To accurately model the CCN formation in exoplanet atmosphere, the nucleation rates of the dominant nucleating species need to be known. Previous studies determined TiO$_2$ \citep{goumans_stardust_2013, lee_dust_2015, boulangier_developing_2019, kohn_dust_2021} as an important nucleating species. Other species which are discussed as nucleating species are Al$_2$O$_3$ \citep{gobrecht_bottom-up_2022}, SiO \citep{gail_primary_1986, lee_dust_2015} and VO (Lecoq-Molinos et al. in prep). In addition to TiO$_2$, we also analysed Al$_2$O$_3$ as a possible nucleating species for clusters up to size $N=10$ \citep{gobrecht_bottom-up_2022}. The results were inconclusive (see Appendix \ref{sec:app_al203}) and therefore, for this study, we decided to focus on TiO$_2$.

    If the maximum cluster size is larger than the smallest thermally stable cluster (N$_\mathrm{max}$ > N$_\star$), the nucleation rate $J_\star(V_l)$ is expected to be independent of the choice of the maximum cluster size. If the maximum cluster size is smaller than N$_\star$, we expect to see different nucleation rates depending on the choice of N$_\mathrm{max}$. Therefore, we test different N$_\mathrm{max}$ (7 $\leq$ N$_\mathrm{max}$ $\leq$ 15) and their impact on the cloud particle number density and nucleation rate. For this section, we set $p_\mathrm{gas}= 0.02$ bar and $T_\mathrm{gas}= 1379$ K. We use the full network as described in Sect.~\ref{sec:res_chem}. The cloud particle number densities and nucleation rates for different N$_\mathrm{max}$ can be seen in Fig. \ref{fig:res_nucpersize}. The predicted number densities for 7 $\leq$ N$_\mathrm{max}$ $\leq$ 15 are all within a factor of 2. Furthermore, the peak in nucleation rate becomes smaller and appears later in time for larger clusters.

    The biggest deviation in predicted cloud particle number density can be seen for N$_\mathrm{max}$ = 13 and N$_\mathrm{max}$ = 15 which predict lower cloud particle number densities than the rest. Looking at the Gibbs free energy per monomer unit ($G^\ominus_\mathrm{TiO_2(N)} / \mathrm{N}$) reveals that these sizes are the only N-mers that have a higher Gibbs free energy per monomer than their (N-1)-mers:
    \begin{align}
        &\frac{G^\ominus_\mathrm{TiO_2(13)}}{\mathrm{13}} - \frac{G^\ominus_\mathrm{TiO_2(12)}}{\mathrm{12}} = 1.158, \\
        &\frac{G^\ominus_\mathrm{TiO_2(15)}}{\mathrm{15}} - \frac{G^\ominus_\mathrm{TiO_2(14)}}{\mathrm{14}} = 0.172.
    \end{align}
    Therefore, N = 13 and N = 15 are less thermally stable than their predecessors. Previous studies have shown the same preference for even N clusters \citep{lasserus_vanadiumv_2019} that we find for the (TiO$_2$)$_N$ clusters but further studies of larger sizes clusters are needed to determine if it is a size dependent trend that can affect the nucleation process. Because our nucleation rate is determined by the largest cluster size, having N$_\mathrm{max}$ = 13 or N$_\mathrm{max}$ = 15 naturally results in lower cloud particle number densities. Thermodynamic data for TiO$_2$ clusters larger than N = 15 is needed to further test this, and to find the thermally stable cluster size $\mathrm{N}_\star$ for TiO$_2$.

    Few studies have already evaluated nucleation using a non-classical approach. \citet{lee_dust_2015} compared the nucleation rate predicted by CNT, MCNT and non-classical nucleation for various temperatures without considering surface growth. We calculated the cloud particle number density over the same temperature range as they analysed (Fig. \ref{fig:res_nucpertemp}). In contrast to our work, they considered only N$_\mathrm{max} = 10$ and only monomer nucleation. Both their and our study predict significant nucleation of TiO$_2$ for temperatures up to roughly 1300 K. Above that, the nucleation of TiO$_2$ quickly decreases. For temperatures between roughly 600 K to 1200 K, our model predicts approximately constant cloud particle number densities whereas the nucleation rate of \citet{lee_dust_2015} decreases. This difference can be traced back to the polymer nucleation. \citet{boulangier_developing_2019} showed that for TiO$_2$ and other nucleation seeds monomer nucleation can underestimate the formation of larger clusters in colder environments.

\section{Time evolution of cloud formation in HD~209458~b}
\label{sec:fully_kinetic}

    \begin{figure}
       \centering
       \includegraphics[width=\hsize]{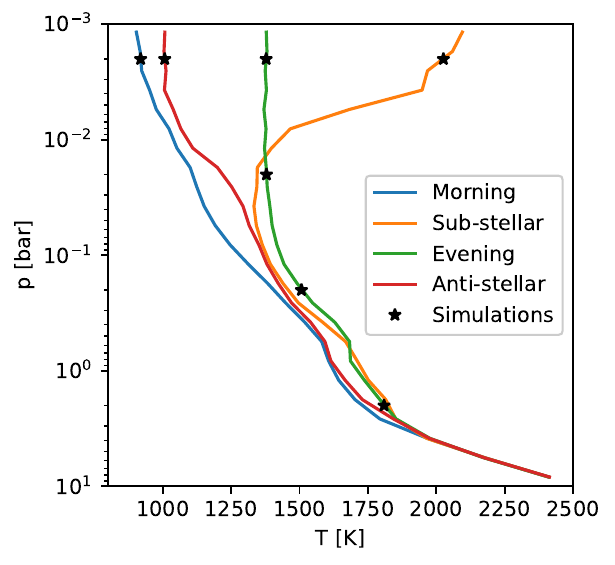}
       \caption{$T_\mathrm{gas}$-$p_\mathrm{gas}$ profiles of HD~209458~b taken from \citet{schneider_exploring_2022}. The $T_\mathrm{gas}$-$p_\mathrm{gas}$ points chosen for our simulations are marked with~$\star$.}
       \label{fig:hd2_profiles}
    \end{figure}

    \begin{figure}
       \centering
       \includegraphics[width=\hsize]{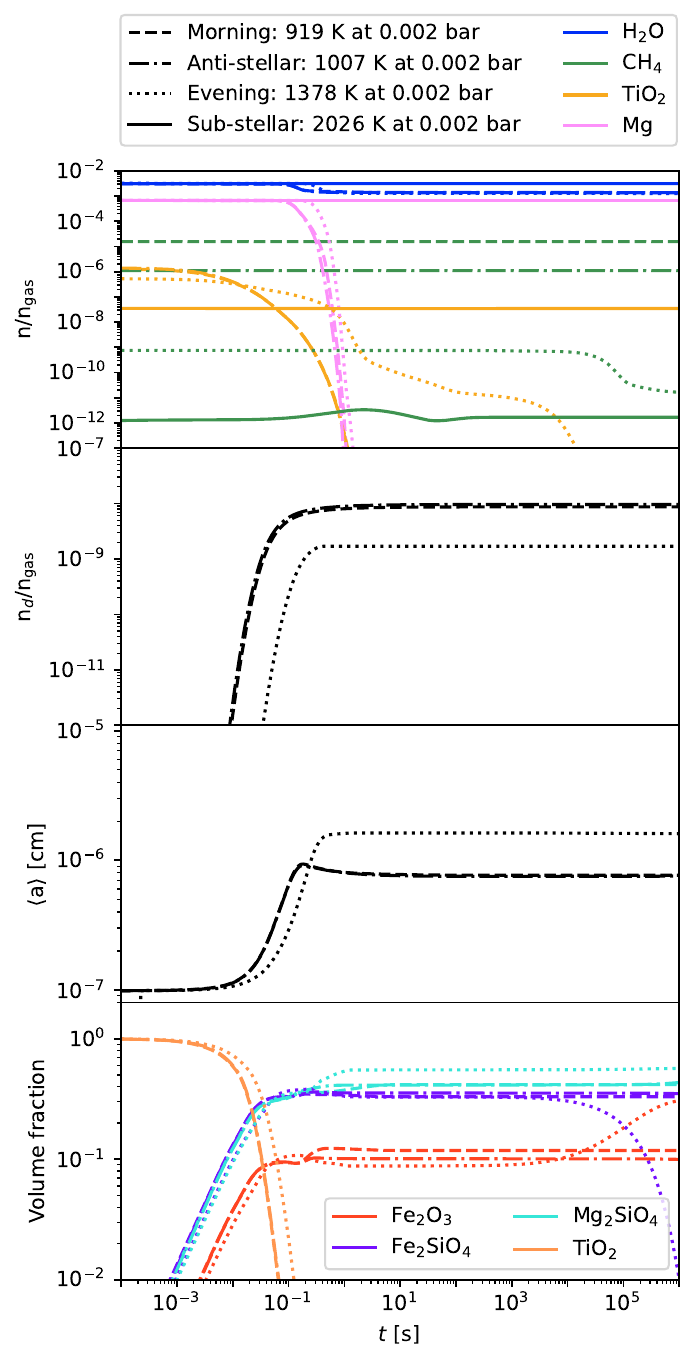}
       \caption{Concentrations of selected gas-phase species (\textbf{Top}), cloud particle number density (\textbf{Upper middle}), mean cloud particle size (\textbf{Lower middle}), and selected volume fractions (\textbf{Bottom}) at the sub-stellar point, evening terminator, anti-stellar point and morning terminator at $p_\mathrm{gas}= 0.002$ bar. The sub-stellar point does not form clouds.}
       \label{fig:hd2_p}
    \end{figure}
    \begin{figure}
       \centering
       \includegraphics[width=\hsize]{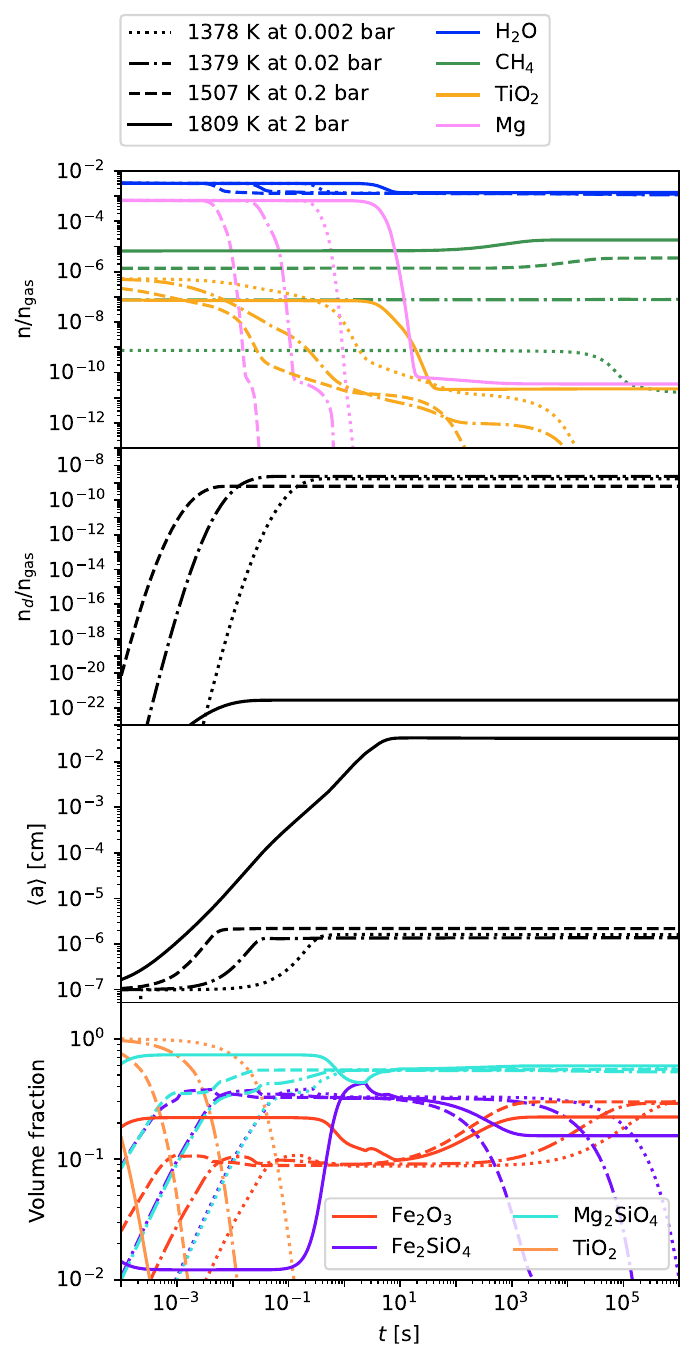}
       \caption{Concentrations of selected gas-phase species (\textbf{Top}), cloud particle number density (\textbf{Upper middle}), mean cloud particle size (\textbf{Lower middle}), and selected volume fractions (\textbf{Bottom}) for logarithmically spaced pressures along the evening terminator.}
       \label{fig:hd2_t}
    \end{figure}

    \begin{figure}
       \centering
       \includegraphics[width=\hsize]{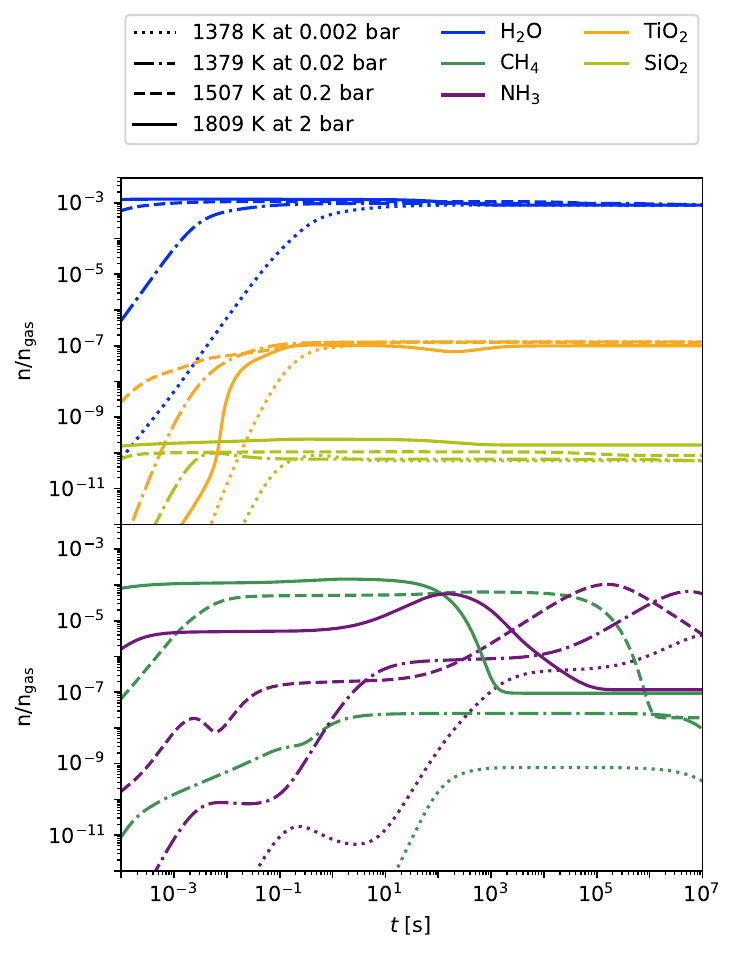}
       \caption{Concentrations of selected gas-phase species along the evening terminator starting from solar-like atomic abundances.}
       \label{fig:atomic}
    \end{figure}

    We simulate the chemistry and cloud formation for various $T_\mathrm{gas}$-$p_\mathrm{gas}$ points within HD~209458~b. The $T_\mathrm{gas}$-$p_\mathrm{gas}$ profiles used in this paper were calculated using expert/MITgcm simulations of HD~209458~b conducted by \citet{schneider_exploring_2022}. The $T_\mathrm{gas}$-$p_\mathrm{gas}$ profiles for the sub-stellar point, anti-stellar point, evening terminator, and morning terminator can be seen in Fig. \ref{fig:hd2_profiles}. The cloud particle concentrations and the mean cloud particle size as well as selected gas-phase concentrations and volume fractions\footnote{All volume fractions can be found in Fig. \ref{fig:hd2_all_volume_frac}.} for the sub-stellar point, evening terminator, anti-stellar point and morning terminator at $p_\mathrm{gas}= 0.002$ bar can be seen in Fig. \ref{fig:hd2_p}. This pressure layer was selected as it showed the largest spread in temperatures. The results for 4 logarithmically spaced pressure points ($p_\mathrm{gas}\in \{0.002 ~\mathrm{bar}, 0.02 ~\mathrm{bar}, 0.2 ~\mathrm{bar}, 2 ~\mathrm{bar}\}$) along the evening terminator can be seen in Fig. \ref{fig:hd2_t}. Only one $T_\mathrm{gas}$-$p_\mathrm{gas}$ profile was selected since the temperatures deeper in the atmosphere ($p_\mathrm{gas} = 0.2$ bar and $p_\mathrm{gas} = 2$ bar) are similar. The evening terminator was selected because of its intermediate temperature in the upper atmosphere at p = 0.002 bar. We use the full network as described in Sect.~\ref{sec:res_chem} for all simulations and start from chemical equilibrium calculated with \texttt{GGchem}.

    To be able to compare gas-phase timescales to cloud formation timescales, we ran the VULCAN+ network (see Sect.~\ref{sec:res_chem}) starting from solar-like atomic abundances for the $T_\mathrm{gas}$-$p_\mathrm{gas}$ points of the evening terminator. The resulting gas-phase concentrations of selected gas-phase species can be seen in Fig. \ref{fig:atomic}. For all $T_\mathrm{gas}$-$p_\mathrm{gas}$ points, the concentrations of H$_2$O, TiO$_2$, and SiO$_2$ quickly approach their stationary values ($\tau_\mathrm{chem} < 1$ s). Their chemical timescale is highly pressure dependent and decreases for higher pressures. CH$_4$ and NH$_3$ on the other hand, still show significant differences for $t > 10^5$ s for all but the highest pressure ($p_\mathrm{gas}= 2$ bar).

    \subsection{Time evolution of cloud formation}
    Our results for the cloud formation within HD 208458 b start from a chemical equilibrium gas-phase from which clouds are formed. These simulations therefore can give us an indication on the timescale of nucleation and bulk growth. The predicted cloud particle concentrations for most $T_\mathrm{gas}$-$p_\mathrm{gas}$ points of HD~209458~b quickly converge to stationary values ($\tau_\mathrm{nuc} < 1$ s). The only exception is the sub-stellar point at $p_\mathrm{gas}= 0.002$ bar where no cloud particles are predicted due to the high temperatures. Comparing the chemical timescales \citep[see Fig. \ref{fig:atomic} or ][]{tsai_toward_2018, mendonca_three-dimensional_2018} to the nucleation timescale shows that nucleation happens on similar timescale as the chemical species with a shorter chemical timescale (e.g. H$_2$O, TiO$_2$, SiO$_2$). The $T_\mathrm{gas}$-$p_\mathrm{gas}$ points at $p_\mathrm{gas}= 0.002$ bar, other than the sub-stellar point, show only a small temperature dependence of the predicted cloud particle concentrations and nucleation timescale. This is not unexpected for temperatures ranging from 919 K to 1378 K. We have shown that TiO$_2$ nucleation is roughly constant for this temperature range (see Sect.~\ref{sec:res_crit}). Similarly, along the evening terminator we see a decrease in cloud particle concentrations consistent with the findings of Sect.~\ref{sec:res_crit}. At $p_\mathrm{gas} = 2$~bar within the evening terminator, the cloud particle concentration reaches only $n_d/n_\mathrm{gas} \approx 10^{-22}$. The lower number of cloud particles results in larger cloud particles because the bulk growth material condense onto fewer particles \citep{helling_exoplanet_2023}. Hence, the average cloud particle size reaches up to 0.033 cm.

    The peak in bulk growth closely follows the peak in nucleation and also approaches stationary values on timescales shorter than 1 second ($\tau_\mathrm{bulk} < 1$ s). The exception to this is the evening terminator at $p_\mathrm{gas} = 2$~bar. The cloud particles still grow at roughly the same speed, but since much more material is available per cloud particle, it takes longer to reach stationary values for the average cloud particle size.

    In all cases, the volume fractions start out TiO$_2$ dominated. After bulk growth starts the cloud particles become considerably heterogeneous. In all cases, Mg$_2$SiO$_4$ becomes the dominant bulk material and hence also the dominant Mg and Si bearing species. Around 0.1 to 1 second, a short increase in the average cloud particle size can be seen for the morning terminator and the anti-stellar point at $p_\mathrm{gas} = 0.002$~bar. This size increase is caused by a temporary increase in SiO[s], SiO$_2$[s] and Fe$_2$SiO$_4$[s] (see also Fig.~\ref{fig:hd2_all_volume_frac}). Without the SiO-SiO$_2$ cycle the temporary peak of Fe$_2$SiO$_4$ still occurs. This temporary increase is likely a result of feedback between cloud formation and disequilibrium chemistry. Because the cloud formation is directly coupled to the gas phase via the reaction supersaturation ratio, temporary changes in the gas-phase chemistry can be caused by cloud formation and vice versa.

    For the dominant Fe bearing species we see a switch from Fe$_2$SiO$_4$ to Fe$_2$O$_3$ for $t > 10^3$ s. These are similar timescales of CH$_4$ and NH$_3$ \citep[see Fig. \ref{fig:atomic} or ][]{tsai_toward_2018, mendonca_three-dimensional_2018}. It is important to note that the change in the dominant Fe bearing species is not related to the SiO-SiO$_2$ cycle. It still occurs even if SiO[s] and SiO$_2$[s] are not considered as bulk growth species. The timescale of the transition is highly pressure dependent and becomes shorter for higher pressures. Furthermore, the change in composition does not significantly affect the cloud particle size. Here it is important to note that our cloud formation formalism does not include any solid-to-solid composition changes. All changes happen via the gas-phase through bulk growth reactions as described in Sect.~\ref{sec:Model_sr}.

    \citet{helling_dust_2006} analysed the timescales of cloud formation with a similar cloud model as used in this work. In contrast to our work, the gas-phase is assumed to be in equilibrium (and depleted by cloud formation), MCNT is used to describe nucleation, and fewer surface reactions are used. The nucleation and bulk growth timescales they find are similar to ours ($\tau_\mathrm{nuc} < 1$ s and $\tau_\mathrm{bulk} < 1$ s). Coupling gas-phase chemistry and cloud formation does not seem to impact these timescales. However, secondary effects like the change in the dominant Fe-bearing species and the SiO-SiO$_2$ cycle only appear once gas-phase chemistry and cloud formation are fully coupled.

    \citet{powell_formation_2018} also analysed the timescales of cloud formation using a diffusive approach. Their rates are limited by the time it takes for the key species to diffuse to the cloud particle. They calculate their timescales as the number density of cloud particles divided by the influx of new cloud particles once a stationary solution has been reached. Consequently, their growth and nucleation timescales for TiO$_2$ are larger than ours ($\tau_\mathrm{nuc} > 10$ s and $\tau_\mathrm{bulk} > 10$ s).

    \subsection{Comparison to dynamical processes}
    To find whether cloud formation happens in disequilibrium or is affected by disequilibrium chemistry, we compare our results to different dynamical processes.

    \subsubsection{Gravitational settling}
    Cloud particles in exoplanet atmospheres gravitationally settle over time. Whether gravitational settling timescales are faster than cloud formation timescales depends on many factors such as the bulk growth speed, bulk growth material replenishment, and the frictional force of cloud particles within the atmosphere \citep{woitke_dust_2003}. For smaller particles ($\langle a \rangle < 10^{-4}$ cm), growth is generally more efficient than gravitational settling \citep{woitke_dust_2003, powell_formation_2018}. However, if the condition favour larger particles and gravitational settling becomes more efficient "cold traps" can occur \citep{parmentier_3d_2013, parmentier_transitions_2016, powell_formation_2018} where most cloud material is concentrated at the cloud base.

    Comparing our cloud particle number densities and average radii to \citet{powell_formation_2018} reveals differences that can be explained by gravitational settling and replenishment. In contrast to our work, they generally predict less and larger cloud particles. Settling removes cloud particles from the atmosphere thus leading to less particles. The replenished material then condenses onto already existing particles. Since fewer particles are present, they become larger. The exception to this is the evening terminator at at $p_\mathrm{gas} = 2$~bar. Here, nucleation is so inefficient that we predict very large cloud particles ($\langle a \rangle = 0.033$ cm). However, these particles would quickly settle down and they are unlikely to persist in a 1D model.

    \subsubsection{Vertical and horizontal transport}
    Quenching occurs when the chemical timescale $\tau_\mathrm{chem}$ [s] becomes larger then the vertical mixing timescale $\tau_\mathrm{dyn}$ [s] \citep{moses_chemical_2014}. In low density environments, chemical timescales typically become longer \citep{tsai_toward_2018} and dynamical timescales typically become smaller \citep{parmentier_3d_2013}. Therefore, quenching becomes more relevant in the upper atmosphere of exoplanets \citep{baeyens_grid_2021}. Typical vertical mixing timescales are between $10^{3}$ s < $\tau_\mathrm{dyn}$ < $10^{7}$ s \citep{agundez_pseudo_2014, drummond_observable_2018, baeyens_grid_2021}. Our results show that nucleation happens on much shorter timescales than this ($\tau_\mathrm{nuc}$ < 1 s) and is therefore expected to be less affected by quenching. The cloud particle composition on the other hand changes on longer timescales which are similar to the chemical timescales of CH$_4$ and NH$_3$. Both CH$_4$ and NH$_3$ are known to be gas-phase species affected by quenching \citep{moses_disequilibrium_2011}. Therefore, the cloud particle composition might be susceptible to quenching as well.

    Similar to the vertical timescale, one can compare the nucleation and bulk growth timescales to the horizontal mixing timescales which consists of the latitudinal timescale and the longitudinal timescale. \citet{mendonca_three-dimensional_2018} analysed\footnote{WASP-43 b ($T_\mathrm{eq} = 1400$ K) is a hot Jupiter with a slightly colder equilibrium temperature than HD~209458~b ($T_\mathrm{eq} = 1500$ K) \citep{helling_cloud_2021}.} WASP-43 b and found that latitudinal mixing happens on similar timescales as the vertical timescale. Longitudinal mixing on the other hand can be orders of magnitude shorter. This is mostly due to the strong equatorial wind jets. In their analysis, all (longitudinal, latitudinal and vertical) mixing timescales are well above $10^{3}$ s > $\tau_\mathrm{dyn}$. Therefore, similar to quenching, nucleation and the peak in bulk growth might be less affected by horizontal mixing. The cloud particle composition on the other hand might be impacted.

    \subsubsection{Stellar Flares}
    If periodic effects disturb the chemical abundances, the relaxation timescale $\tau_\mathrm{relax}$ indicates how quickly the chemical abundances return back to pre-disruption values. An example for an effect that temporarily impacts chemistry are stellar flares. They periodically enhance the radiation received by a planet and cause chemical disequilibrium through photochemistry \citep{tilley_modeling_2019, louca_impact_2022}. The chemical relaxation timescale after a stellar flare event can be on the order of hours \citep[$\tau_\mathrm{relax}$ > $10^3$~s;][]{konings_impact_2022}. In the chemical relaxation scheme \citep{smith_estimation_1998, cooper_dynamics_2006, kawashima_implementation_2021} the relaxation timescale is given by the chemical timescale. This scheme typically also finds relaxation timescales of $\tau_\mathrm{relax}$ > $10^3$~s for CH$_4$ and NH$_3$ \citep[see Fig. \ref{fig:atomic} or ][]{tsai_toward_2018, mendonca_three-dimensional_2018}. Nucleation and the peak in bulk growth occur on much shorter timescales than this and therefore can adjust to the temporary chemical disequilibrium. The cloud particle composition on the other hand takes longer to transition and might not adjust to the temporary chemical disequilibrium.

\section{Summary}
\label{sec:Conclusion}

    We established a fully kinetic cloud formation description combining disequilibrium chemistry, kinetic nucleation and bulk growth through condensation and surface reactions. The kinetic gas-phase chemistry network for this study was based on the "NCHO thermo network" of VULCAN. This network was expanded with Ti (and Si) reactions to connect the gas-phase chemistry with the kinetic nucleation of TiO$_2$. We considered TiO$_2$ polymer nucleation using cluster data up to cluster size N = 15. For the bulk material, we considered TiO$_2$[s], Al$_2$O$_3$[s], SiO[s], SiO$_2$[s], Mg$_2$SiO$_4$[s], FeO[s], FeS[s] Fe$_2$O$_3$[s], and Fe$_2$SiO$_4$[s]. These materials can grow through 59 bulk growth reactions.

    The kinetic polymer nucleation of TiO$_2$ indicates a similar temperature and pressure dependence as previous non-classical studies. We tested different maximum cluster sizes between N$_\mathrm{max} = 7$ and N$_\mathrm{max} = 15$ and found differences of the predicted cloud particle number density within a factor of 2. For further investigations, thermodynamic data of larger TiO$_2$ clusters is required. Overall, our results suggest that kinetic nucleation is a viable alternative to classical nucleation theory if the cluster data of the nucleating species is available. The required cluster sizes depend on the nucleating species.

    The fully kinetic description of surface reactions resulted in a SiO-SiO$_2$ catalytic cycle that dissociates H$_2$ into 2 H. The increase in atomic hydrogen can reduce the CH$_4$ abundance by over an order of magnitude. If this catalytic cycle occurs in exoplanet atmospheres remains to be seen.

    We simulated the chemistry of various $T_\mathrm{gas}$-$p_\mathrm{gas}$ points within the atmosphere of HD~209458~b. For all $T_\mathrm{gas}$-$p_\mathrm{gas}$ points, except the sub-stellar point, nucleation and bulk growth occurred. In all cases were nucleation occurred, nucleation and bulk growth reached a stationary behavior within 1 s. A comparison to the timescales of quenching and chemical relaxation showed that nucleation can happen on much shorter timescales. Hence, our work confirms that the assumption of localised nucleation is generally justified in exoplanet atmospheres. For the cloud particle composition on the other hand, we found changes on the timescale larger than 10$^{3}$ seconds. This indicates that the composition of cloud particles can be susceptible to quenching.

\begin{acknowledgements}
S.K, H.LM., Ch.H., and L.D. acknowledge funding from the European Union H2020-MSCA-ITN-2019 under grant agreement no. 860470 (CHAMELEON). N.B acknowledges financial support from the Austrian Academy of Sciences.
\end{acknowledgements}

\bibliographystyle{aa} 
\bibliography{bibliography_ch} 

\begin{appendix}

    \section{Numerical solver}
    \label{sec:Model_solver}

    The chemical kinetic network can be described as an ordinary differential equation (ODE). Therefore, a fast and accurate ODE solver is required. Similar to other works \citep[e.g.][]{agundez_pseudo_2014, rimmer_chemical_2016, molaverdikhani_cold_2019, baeyens_grid_2021}, we chose DLSODES from ODEPACK \citep{hindmarsh_odepack_1983, radhakrishnan_description_1993}.

\begin{figure}
       \centering
       \includegraphics[width=\hsize]{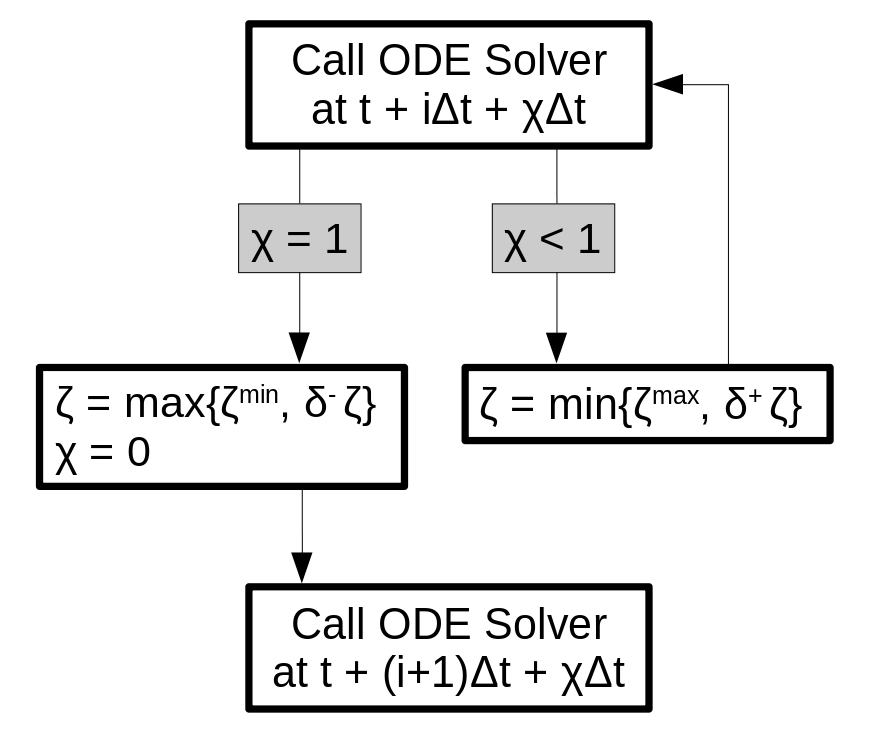}
       \caption{Dynamic tolerances scheme for the ODE solver.}
       \label{fig:dyntol}
    \end{figure}

    Combining chemistry, nucleation and bulk growth through condensation and surface reactions results in a stiff ODE system, especially during the onset of nucleation and the onset of bulk growth. To prevent the solver from getting stuck, we introduce dynamic tolerances. During the iteration, the solver tries to compute a given time step $\Delta t$ with a given relative tolerance $\zeta_\mathrm{rel}$ and a given absolute tolerance $\zeta_\mathrm{abs}$. If the solver only reaches a percentage $\chi < 1$ of the time-step, $\zeta_\mathrm{rel}$ and $\zeta_\mathrm{abs}$ are increased by a factor of $\delta^+_\mathrm{rel}$ and $\delta^+_\mathrm{abs}$, respectively. If the solver reaches the full time step $\chi = 1$, the relative and absolute tolerances are decreased by a factor of $\delta^-_\mathrm{rel}$ and $\delta^-_\mathrm{abs}$, respectively. To ensure sufficient precision of the ODE, the tolerances have an upper limit $\zeta^\mathrm{max}_\mathrm{rel}$ and $\zeta^\mathrm{max}_\mathrm{abs}$, and a lower limit $\zeta^\mathrm{min}_\mathrm{rel}$ and $\zeta^\mathrm{min}_\mathrm{abs}$, respectively. The schematics of the algorithm is show in Fig. \ref{fig:dyntol}. In this paper, the following values are used:
    \begin{align}
        \zeta^\mathrm{max}_\mathrm{rel} &= 10^{-3} \\
        \zeta^\mathrm{max}_\mathrm{abs} &= 10^{-40} \\
        \zeta^\mathrm{min}_\mathrm{rel} &= 10^{-9} \\
        \zeta^\mathrm{min}_\mathrm{abs} &= 10^{-60} \\
        \delta^-_\mathrm{rel} &= \delta^-_\mathrm{abs} = 0.1 \\
        \delta^+_\mathrm{rel} &= \delta^+_\mathrm{abs} = 1.1
    \end{align}
    The starting tolerance is always equal to the minimum tolerance and was selected depending on the simulation.

    \section{GGchem}
    \label{sec:app_ggchem}

    We consider the following species in our chemical equilibrium calculations done with \texttt{GGchem}:
    OH, H$_2$, H$_2$O, H, O, CH, C, CH$_2$, CH$_3$, CH$_4$, C$_2$, C$_2$H$_2$, C$_2$H, C$_2$H$_3$, C$_2$H$_4$, C$_2$H$_5$, C$_2$H$_6$, CO, CO$_2$, CH$_2$OH, H$_2$CO, HCO, CH$_3$O, CH$_3$OH, CH$_3$CO, O$_2$, H$_2$CCO, HCCO, N, NH, CN, HCN, NO, NH$_2$, N$_2$, NH$_3$, N$_2$H$_2$, N$_2$H, N$_2$H$_3$, N$_2$H$_4$, HNO, H$_2$CN, HNCO, NO$_2$, N$_2$O, C$_4$H$_2$, CH$_2$NH$_2$, CH$_2$NH, CH$_3$NH$_2$, CH$_3$CHO, HNO$_2$, NCO, HO$_2$, H$_2$O$_2$, HC$_3$N, CH$_3$CN, CH$_2$CN, C$_2$H$_3$CN, C$_3$H$_3$, C$_3$H$_2$, C$_3$H$_4$, C$_4$H$_3$, C$_4$H$_5$, C$_6$H$_6$, C$_6$H$_5$, (O)$_1$, (CH$_2$)$_1$, (N$_2$)$_D$, He, Ti, TiO, SO$_2$, SO, TiO$_2$, SiO$^+$, Si$^+$, SiH$^+$, H$^-$, Si, HF, SiF$^+$, He$^+$, SiO$_2$, Na, Na$^+$, SiO, HCO$^+$, P$^+$, P, S$^+$, S, e$^-$, Fe, Fe$^+$, F, (TiO$_2$)$_2$, (TiO$_2$)$_3$, (TiO$_2$)$_4$, (TiO$_2$)$_5$, (TiO$_2$)$_6$, (TiO$_2$)$_7$, (TiO$_2$)$_8$, (TiO$_2$)$_9$, (TiO$_2$)$_{10}$, (TiO$_2$)$_{11}$, (TiO$_2$)$_{12}$, (TiO$_2$)$_{13}$, (TiO$_2$)$_{14}$, (TiO$_2$)$_{15}$, Mg, Al, Ca, Cl, FeO$_2$, Al$_2$O$_3$, AlH$_2$, AlH$_3$, Al(OH)$_2$, Al(OH)$_3$, H$_2$S$_2$, F$_2$, Mg$_2$, Al$_2$, Si$_2$, S$_2$, Cl$_2$, MgH, AlH, SiH, HS, HCl, CaH, TiH, FeH, CF, SiC, CS, CCl, FN, AlN, SiN, SN, NCl, TiN, FO, MgO, AlO, ClO, CaO, FeO, MgF, AlF, SiF, SF, CaF, MgS, AlS, SiS, CaS, TiS, MgCl, AlCl, SiCl, CaCl, FeCl, AlClF, AlClF$_2$, AlOCl, AlCl$_2$, AlCl$_2$F, AlCl$_3$, AlOF, AlF$_2$, AlF$_2$O, AlF$_3$, AlOH, HAlO, AlO$_2$H, AlO$_2$, Al$_2$Cl$_6$, (AlF$_3$)$_2$, Al$_2$O, Al$_2$O$_2$, AlC, CFClO, CClF$_3$, CClN, CClO, CCl$_2$, CCl$_2$F$_2$, COCl$_2$, CCl$_3$, CCl$_3$F, CCl$_4$, CFN, CFO, CF$_2$, CF$_2$O, CF$_3$, CF$_4$, CF$_4$O, CF$_8$S, CHCl, CHCl$_3$, CHF, CHFO, CHF$_3$, CHNO, CH$_2$Cl$_2$, CH$_2$ClF, CH$_2$F$_2$, CH$_3$Cl, CH$_3$F, CNO, CNN, NCN, COS, CS$_2$, Si$_2$C, C$_2$Cl$_2$, C$_2$Cl$_4$, C$_2$Cl$_6$, C$_2$F$_2$, C$_2$F$_3$N, C$_2$F$_4$, C$_2$F$_6$, C$_2$HCl, C$_2$HF, C$_2$H$_4$O, C$_2$N, C$_2$N$_2$, SiC$_2$, C$_2$O, C$_3$, C$_3$O$_2$, C$_4$, C$_4$N$_2$, C$_5$, Fe(CO)$_5$, CaCl$_2$, CaF$_2$, CaOH, Ca(OH)$_2$, Ca$_2$, ClF, MgClF, ClFO$_2$S, ClFO$_3$, ClF$_3$, ClF$_3$Si, ClF$_5$, ClF$_5$S, CHClF$_2$, CHCl$_2$F, OHCl, SiH$_3$Cl, NOCl, NO$_2$Cl, TiOCl, ClO$_2$, ClO$_3$, SCl, ClS$_2$, TiCl, FeCl$_2$, SiH$_2$Cl$_2$, MgCl$_2$, ClOCl, ClClO, TiOCl$_2$, ClO$_2$Cl, ClOClO, SO$_2$Cl$_2$, SCl$_2$, SiCl$_2$, TiCl$_2$, SiFCl$_3$, FeCl$_3$, SiHCl$_3$, SiCl$_3$, TiCl$_3$, Fe$_2$Cl$_4$, Mg$_2$Cl$_4$, SiCl$_4$, TiCl$_4$, (FeCl$_3$)$_2$, FeF, FHO, FHO$_3$S, SiH$_3$F, FNO, FNO$_2$, FNO$_3$, TiOF, OFO, FOO, TiF, FeF$_2$, H$_2$F$_2$, SiH$_2$F$_2$, MgF$_2$, F$_2$N, F$_2$N$_2$(cis), F$_2$N$_2$(trans), F$_2$O, F$_2$OS, SiOF$_2$, TiOF$_2$, F$_2$O$_2$, F$_{2}$O$_{2}$S, SF$_2$, FSSF, F$_2$S$_2$, SiF$_2$, TiF$_2$, FeF$_3$, SiHF$_3$, F$_3$H$_3$, NF$_3$, NOF$_3$, SF$_3$, SiF$_3$, TiF$_3$, F$_4$H$_4$, Mg$_2$F$_4$, N$_2$F$_4$, SF$_4$, SiF$_4$, TiF$_4$, F$_5$H$_5$, SF$_5$, F$_6$H$_6$, SF$_6$, F$_7$H$_7$, F$_{10}$S$_2$, Fe(OH)$_2$, FeS, MgOH, HONO, HNO$_3$, Mg(OH)$_2$, H$_2$SO$_4$, H$_2$S, SiH$_4$, MgN, NO$_3$, Si$_2$N, N$_2$O$_3$, N$_2$O$_4$, N$_2$O$_5$, N$_3$, S$_2$O, O$_3$, SO$_3$, S$_3$, S$_4$, S$_5$, S$_6$, S$_7$, S$_8$, Si$_3$, TiC, Si(CH$_3$)$_4$, SiCH$_3$Cl$_3$, SiH$_2$, SiH$_3$, TiC$_2$, C$_3$H, Si$_2$C$_2$, and TiC$_4$.

    \section{\texorpdfstring{Al$_2$O$_3$}{Lg} nucleation}
    \label{sec:app_al203}

    To calculate the nucleation rate of potential nucleation species, thermodynamic data of their clusters needs to be available. With TiO$_2$, we are in the fortunate position of having data available up to N = 15 \citep{sindel_revisiting_2022}. For Al$_2$O$_3$ we only have data up to N = 10 from \citet{gobrecht_bottom-up_2022}. They also derived a detailed chemical kinetics network up to (Al$_2$O$_3$)$_4$, including a small C-H-O network for the gas-phase.

    Similar as in Sect.~\ref{sec:res_crit}, we calculated the predicted cloud particle number densities and nucleation rate of Al$_2$O$_3$ using 5 $\leq$ N$_\mathrm{max}$ $\leq$ 10, but the results where inconclusive and no stable nucleation behaviour was found. Neither the cloud particle number density nor the nucleation rate showed any clear trends with N$_\mathrm{max}$. A possible explanation for this non-convergence could be that the thermally stable cluster is (much) larger than N$_\mathrm{max}$ = 10 or that Al$_2$O$_3$ does not follow homogeneous nucleation. More studies on the formation of Al$_2$O$_3$ clusters is required to model the kinetic nucleation of Al$_2$O$_3$.

    \section{Comparing chemical networks}
    \label{sec:app_chemical_networks}

    The cloud formation model described in this paper can be combined with any chemical kinetics network. Typically, the more extensive the network, the longer individual model runs take. In the development of this model, we first relayed on a small C-H-O network taken from \citet{gobrecht_bottom-up_2022} (\textbf{Gobrecht}). Later we used the chemical network of \citet{tsai_vulcan_2017} (\textbf{Vulcan}). Here, we compare the two networks.

    Throughout this section, we reference multiple reactions from the chemical network of \textbf{Gobrecht} and use the following short hand:
    \begin{itemize}
        \item (G02): H$_2$ + H$_2$ $\rightarrow$ H + H + H$_2$
        \item (G04): H$_2$ + H $\rightarrow$ H + H + H
        \item (G06): H$_2$ + He $\rightarrow$ H + H + He
        \item (G11): OH + H + H$_2$O $\rightarrow$ H$_2$O + H$_2$O
        \item (G12): H$_2$O + H$_2$O $\rightarrow$ OH + H + H$_2$O
        \item (G24): OH + H + M $\rightarrow$ H$_2$O + M
        \item (G25): H$_2$O + M $\rightarrow$ OH + H + M
    \end{itemize}

    \subsection{Gobrecht vs Vulcan}
    \label{sec:App_gob_vs_vuc}

\begin{figure}
       \centering
       \includegraphics[width=\hsize]{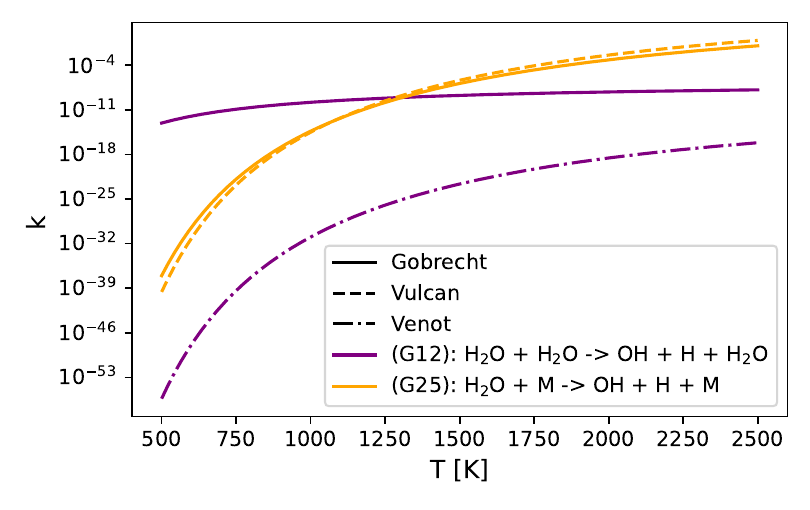}
       \caption{Comparison of reaction rate coefficients at $p_\mathrm{gas} = 0.01$ bar}. The unit of k is [cm$^3$~s$^{-1}$] for G12 and [s$^{-1}$] for G25.
       \label{fig:reac_test}
    \end{figure}

    With three exceptions, all reactions of \textbf{Gobrecht} can be found within \textbf{Vulcan}. The first exceptions are reaction G11 and G12, which do not appear in \textbf{Vulcan}. Especially reaction G12 has a significant impact on the concentration of all species. The reaction rate for G12, and its corresponding forward reaction G11, were derived in supersonic combustion chemistry experiments at 2790~K~<~T~<~3200~K and p~=~250 kPa \citep{javoy_elementary_2003}. A comparison with other chemical networks for exoplanet atmospheres revealed inconsistencies. Instead of G11 and G12, \textbf{Vulcan} only considers the reactions G24 and G25. These are similar reactions but with an arbitrary third body instead of H$_2$O. \citet{venot_chemical_2012} include reactions G11 and G12 but do not include G24 and G25. \citet{rimmer_chemical_2016} do not consider any of the reactions G11, G12, G24, and G25. A comparison between the reaction rate coefficients of G12, and G25 can be seen in Fig. \ref{fig:reac_test}.

    The second difference between \textbf{Gobrecht} and \textbf{Vulcan} is the reaction G17. This reaction describes the radiative association of CO. Since we don't include any photochemical reactions in \textbf{Vulcan}, we also exclude this reaction for \textbf{Gobrecht}. From here on we call the collection of reactions corresponding to the network of \textbf{Gobrecht} without reactions G11, G12 and G17 the 'base' network.

    The third difference is the description of H+H reactions. In \textbf{Gobrecht}, they are described as 3-body reactions with specific collision partners H, H$_2$ and He (G02, G04, and G06). In \textbf{Vulcan}, these reactions are described with a generic 3rd body M. We use each network's description of H+H reactions and don't adjust them.

\begin{figure}
       \centering
       \includegraphics[width=\hsize]{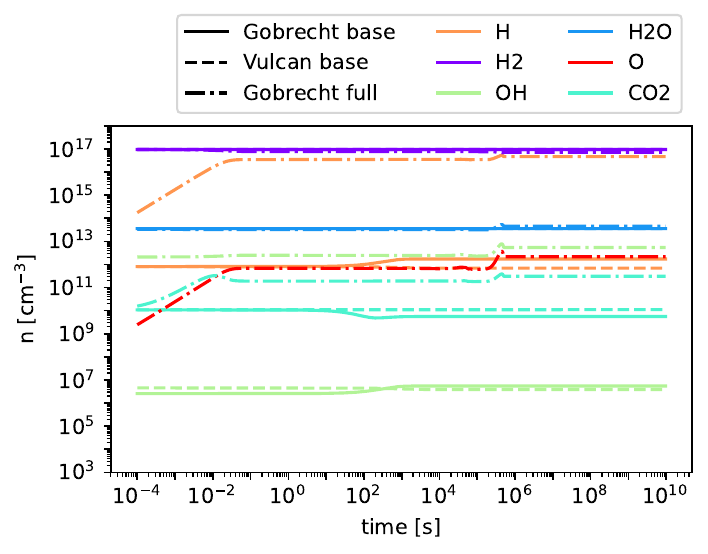}
       \caption{Comparison of the chemical network of \textbf{Gobrecht} and \textbf{Vulcan}.}
       \label{fig:gob_vs_vuc}
    \end{figure}

    Evaluating the base network for T = 1300 K and p = 2000 (Fig. \ref{fig:gob_vs_vuc}) once with reaction rates from \textbf{Gobrecht} and once with reaction rates from \textbf{Vulcan} only showed minor differences in CO$_2$, OH, and H abundance. The differences are within expected offsets due to slightly different reaction rates and different Gibbs free energy values. The full network of \textbf{Gobrecht} still produced significantly different abundances.

    \subsection{Vulcan}

    We separate Vulcan's 780 reactions into three categories:
    \begin{itemize}
        \item \textbf{base:} The base \textbf{Vulcan} network as described in Appendix \ref{sec:App_gob_vs_vuc}.

        \item \textbf{won:} The reactions of the "NCHO thermo network" of \textbf{Vulcan} without Nitrogen species as reactant or product.

        \item \textbf{full:} The full "NCHO thermo network" of \textbf{Vulcan}.
    \end{itemize}

    Comparing the base and won network for $p_\mathrm{gas}= 0.002$ bar and T $\in$ \{500 K, 1300 K, 2000 K\} show little to no differences in H, O, OH, H$_2$, H$_2$O, CO, and CO$_2$ number densities. Using the evaluation from Eq. \ref{eq:chem_compare} we find $P(A, \mathrm{base}, \mathrm{won}) < 10^{-3}$ for all of these species. Furthermore, also the deviations from chemical equilibrium as calculated with \texttt{GGchem} are minimal. Comparing the won and full network for $p_\mathrm{gas}= 0.002$ bar and T $\in$ \{500 K, 1300 K, 2000 K\} showed little to no differences in H, O, OH, H$_2$, H$_2$O, CO, and CO$_2$ abundance. Using the evaluation from Eq. \ref{eq:chem_compare} we find $P(A, \mathrm{won}, \mathrm{full}) < 10^{-4}$ for all of these species.

    Our results suggest that the VULCAN network could be used in the base or won configuration as well if a smaller network is desired. Especially the won network might prove useful since it includes the major species H$_2$O, CO, CO$_2$, and CH$_4$ while having only 416 of the 780 reactions of the "NCHO thermo network" of \textbf{Vulcan}.

    \section{Data and additional plots}

    The Ti and Si reactions selected from \citet{boulangier_developing_2019} and \citet{kiefer_effect_2023} are listed in Table \ref{tab:appendix_chemicalNetwork_nuccond}. Some backward reactions are derived using detailed balance with the following coefficients:
    \begin{align}
        \mathrm{EQS_{18}} &= \exp \left(\frac{G^\ominus_\mathrm{TiO_2} + G^\ominus_\mathrm{H} - G^\ominus_\mathrm{TiO} - G^\ominus_\mathrm{OH}}{k_B T_\mathrm{gas}} \right) \\
        \mathrm{EQS_{20}} &= \left( \frac{k_B T}{p^\ominus} \right) \exp \left(\frac{G^\ominus_\mathrm{TiO_2} + G^\ominus_\mathrm{TiO_2} - G^\ominus_\mathrm{(TiO_2)_2} }{k_B T_\mathrm{gas}}\right) \\
        \mathrm{EQS_{22}} &= \left( \frac{k_B T}{p^\ominus} \right) \exp \left(\frac{G^\ominus_\mathrm{(TiO_2)_2} + G^\ominus_\mathrm{TiO_2} - G^\ominus_\mathrm{(TiO_2)_3} }{k_B T_\mathrm{gas}}\right) \\
        \mathrm{EQS_{24}} &= \left( \frac{k_B T}{p^\ominus} \right) \exp \left(\frac{G^\ominus_\mathrm{(TiO_2)_3} + G^\ominus_\mathrm{TiO_2} - G^\ominus_\mathrm{(TiO_2)_4} }{k_B T_\mathrm{gas}}\right) \\
        \mathrm{EQS_{26}} &= \left( \frac{k_B T}{p^\ominus} \right) \exp \left(\frac{G^\ominus_\mathrm{(TiO_2)_2} + G^\ominus_\mathrm{(TiO_2)_2} - G^\ominus_\mathrm{(TiO_2)_4} }{k_B T_\mathrm{gas}}\right)
    \end{align}

\begin{figure}
       \centering
       \includegraphics[width=\hsize]{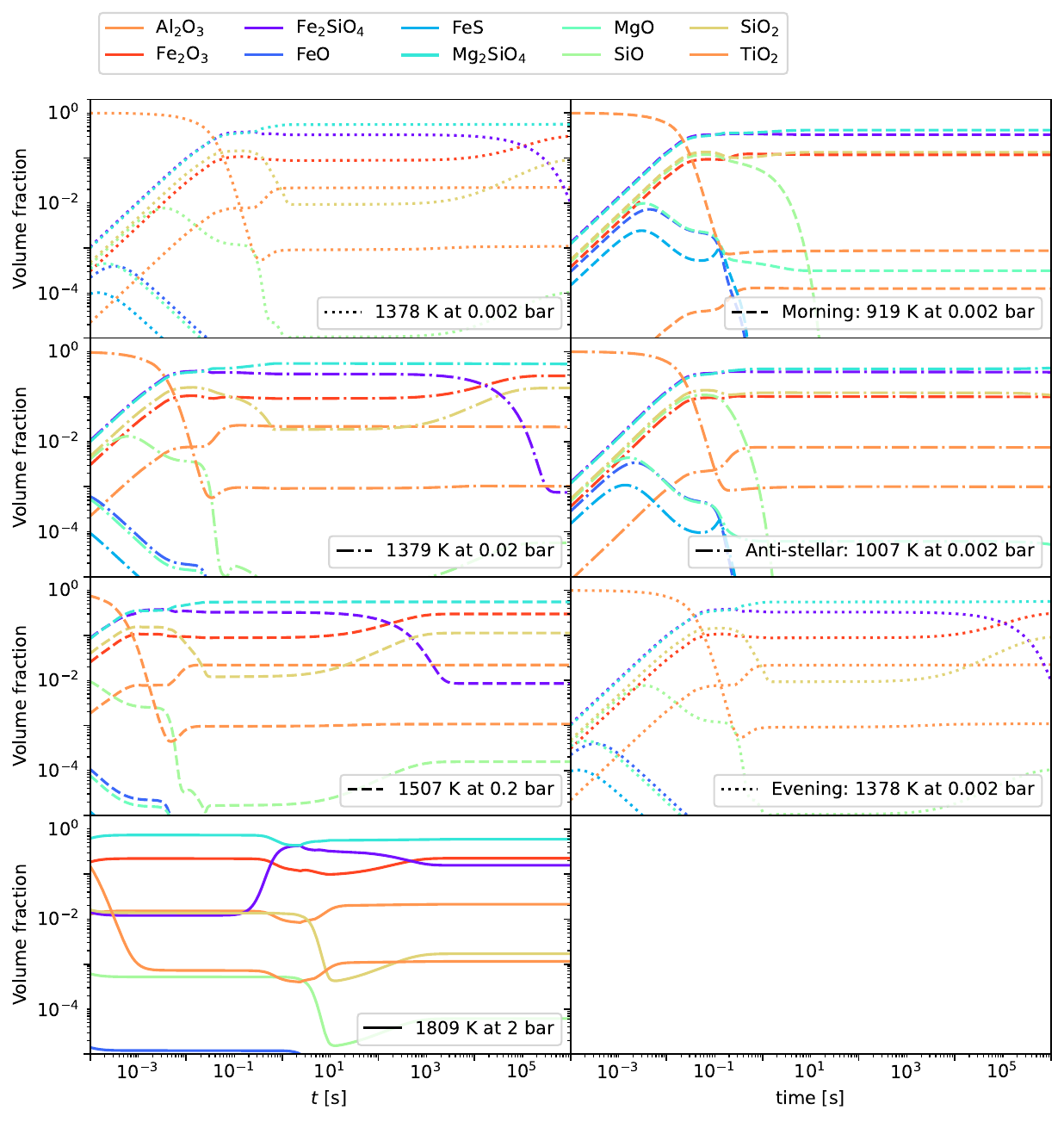}
       \caption{Volume fractions for the $T_\mathrm{gas}$-$p_\mathrm{gas}$ points of HD~209458~b. The sub-stellar point ($p_\mathrm{gas}= 0.002$ bar, $T_\mathrm{gas}= 2026$ K) is not shown since no cloud formation occures.}
       \label{fig:hd2_all_volume_frac}
    \end{figure}

    The condensation reactions taken from \citet{helling_sparkling_2019} and the fitting values for the reaction vapor coefficient (Eq. \ref{eq:s_fit}) are listed in Table \ref{tab:con_reac}. In Fig. \ref{fig:hd2_all_volume_frac}, the volume fraction of all species from the simulations of Sect.~\ref{sec:fully_kinetic} are shown.

    \begin{table*}
        \centering
        \caption{Chemical reactions from \citet{boulangier_developing_2019} and \citet{kiefer_effect_2023} added to the "NCHO thermo network" of VULCAN. }
        \label{tab:appendix_chemicalNetwork_nuccond}
        \begin{tabular}{l l l l}
            \hline\hline
             Index (RNr) & Reaction & Reaction rate coefficient k & Origin \\
             \hline
              1  & Ti + CO$_2$ $\rightarrow$ TiO + CO    & $7       \times 10^{-11} \exp(-14.9/(RT))$             & (1) \\
              2  & TiO + CO $\rightarrow$ Ti + CO$_2$    & $7       \times 10^{-11} \exp(-14.9/(RT))$ EQR$_{564}$ & (1) \\
              3  & Ti + N$_2$O $\rightarrow$ TiO + N$_2$ & $1.74    \times 10^{-10} \exp(-14.3/(RT))$             & (1) \\
              4  & TiO + N2 $\rightarrow$ Ti + N$_2$O    & $1.74    \times 10^{-10} \exp(-14.3/(RT))$ EQR$_{566}$ & (1) \\
              5  & Ti + NO $\rightarrow$ TiO + N         & $3.28    \times 10^{-11} \exp(-3.62/(RT))$             & (1) \\
              6  & TiO + N $\rightarrow$ Ti + NO         & $3.28    \times 10^{-11} \exp(-3.62/(RT))$ EQR$_{565}$ & (1) \\
              7  & Ti + NO$_2$ $\rightarrow$ TiO + NO    & $9       \times 10^{-11}                 $             & (1) \\
              8  & TiO + NO $\rightarrow$ Ti + NO$_2$    & $9       \times 10^{-11}                 $ EQR$_{567}$ & (1) \\
              9  & Ti + O$_2$ $\rightarrow$ TiO + O      & $1.69    \times 10^{-10} \exp(-11.6/(RT))$             & (1) \\
              10 & TiO + O $\rightarrow$ Ti + O$_2$      & $1.69    \times 10^{-10} \exp(-11.6/(RT))$ EQR$_{569}$ & (1) \\
              11 & Ti + SO$_2$ $\rightarrow$ TiO + SO    & $1.7     \times 10^{-10} \exp(-2.66/(RT))$             & (1) \\
              12 & TiO + SO $\rightarrow$ Ti + SO$_2$    & $1.7     \times 10^{-10} \exp(-2.66/(RT))$ EQR$_{572}$ & (1) \\
              13 & TiO + NO $\rightarrow$ TiO$_2$ + N    & $2.2     \times 10^{-12}$                              & (4) \\
              14 & TiO$_2$ + N $\rightarrow$ TiO + NO    & $2.2     \times 10^{-12}$                  EQR$_{574}$ & (4) \\
              15 & TiO + O$_2$ $\rightarrow$ TiO$_2$ + O & $7.07    \times 10^{-12}$                              & (5) \\
              16 & TiO$_2$ + O $\rightarrow$ TiO + O$_2$ & $7.07    \times 10^{-12}$                  EQR$_{570}$ & (5) \\
              17 & TiO + OH $\rightarrow$ TiO$_2$ + H    & $2.07    \times 10^{-10}$ T$^{0.39}$                   & (3) \\
              18 & TiO$_2$ + H $\rightarrow$ TiO + OH    & $2.07    \times 10^{-10}$ T$^{0.39}$       EQS$_{18}$  & (0) \\
              19 & (TiO$_2$)$_2$ + M $\rightarrow$ TiO$_2$ + TiO$_2$ + M             & $1.4 \times 10^{-4} \exp(-48870/T)$                & (3) \\
              20 & TiO$_2$ + TiO$_2$ + M $\rightarrow$ (TiO$_2$)$_2$ + M             & $1.4 \times 10^{-4} \exp(-48870/T)$   EQS$_{20}$   & (6) \\
              21 & (TiO$_2$)$_3$ + M $\rightarrow$ (TiO$_2$)$_2$ + TiO$_2$ + M       & $1.4 \times 10^{-9} \exp(-62411/T)$                & (6) \\
              22 & (TiO$_2$)$_2$ + TiO$_2$ + M $\rightarrow$ (TiO$_2$)$_3$ + M       & $1.4 \times 10^{-9} \exp(-62411/T)$   EQS$_{22}$   & (6) \\
              23 & (TiO$_2$)$_4$ + M $\rightarrow$ (TiO$_2$)$_3$ + TiO$_2$ + M       & $1.4 \times 10^{-9} \exp(-53569/T)$                & (6) \\
              24 & (TiO$_2$)$_3$ + TiO$_2$ + M $\rightarrow$ (TiO$_2$)$_4$ + M       & $1.4 \times 10^{-9} \exp(-53569/T)$   EQS$_{24}$   & (6) \\
              25 & (TiO$_2$)$_4$ + M $\rightarrow$ (TiO$_2$)$_2$ + (TiO$_2$)$_2$ + M & $1.4 \times 10^{-9} \exp(-57194/T)$                & (6) \\
              26 & (TiO$_2$)$_2$ + (TiO$_2$)$_2$ + M $\rightarrow$ (TiO$_2$)$_4$ + M & $1.4 \times 10^{-9} \exp(-57914/T)$   EQS$_{26}$   & (6) \\

              \hline
              27 & Si + OH $\rightarrow$ SiO + H              & $10^{-10}$                                                             & (2) \\
              28 & Si + CO $\rightarrow$ SiO + C              & $1.3 \times 10^{-9} \exp(-34513/\mathrm{T})$                           & (2) \\
              29 & Si + CO$_2$ $\rightarrow$ SiO + CO         & $2.72 \times 10^{-11} \exp(-282/\mathrm{T})$                           & (2) \\
              30 & Si + NO $\rightarrow$ SiO + N              & $9 \times 10^{-11} (\mathrm{T}/300)^{-0.96} \exp(-28/\mathrm{T})$      & (2) \\
              31 & Si + O$_2$ $\rightarrow$ SiO + O           & $1.72 \times 10^{-10} (\mathrm{T}/300)^{-0.53} ~ \exp(-17/\mathrm{T})$ & (2) \\
              32 & Si + O $\rightarrow$ SiO                   & $5.52 \times 10^{-18} (\mathrm{T}/300)^{0.31}$                         & (2) \\
              33 & Si + OH $\rightarrow$ SiO + H              & $1 \times 10^{-10}$                                                    & (2) \\
              34 & SiO + OH $\rightarrow$ SiO$_2$ + H         & $2 \times 10^{-12}$                                                    & (2) \\
              35 & SiO$_2$ + H $\rightarrow$ SiO + OH         & $2 \times 10^{-12}$ EQR$_{516}$                                        & (2) \\
              36 & Si + HCO$^+$ $\rightarrow$ SiH$^+$ + CO    & $1.6 \times 10^{-9}$                                                   & (2) \\
              37 & Si + P$^+$ $\rightarrow$ Si$^+$ + P        & $1 \times 10^{-9}$                                                     & (2) \\
              38 & Si + He$^+$ $\rightarrow$ Si$^+$ + He      & $3.3 \times 10^{-9}$                                                   & (2) \\
              39 & Si + S$^+$ $\rightarrow$ Si$^+$ + S        & $1.6 \times 10^{-9}$                                                   & (2) \\
              40 & Si$^+$ + OH $\rightarrow$ SiO$^+$ + H      & $6.3 \times 10^{-10} (\mathrm{T}/300)^{-0.5}$                          & (2) \\
              41 & Si$^+$ + H $\rightarrow$ Si + H            & $1.17 \times 10^{-17} (\mathrm{T}/300)^{-0.14}$                        & (2) \\
              42 & Si$^+$ + H$^-$ $\rightarrow$ Si + H        & $7.51 \times 10^{-8} (\mathrm{T}/300)^{-0.5}$                          & (2) \\
              43 & Si$^+$ + HF $\rightarrow$ SiF$^+$ + H      & $5.7 \times 10^{-9} (\mathrm{T}/300)^{-0.5}$                           & (2) \\
              44 & Si$^+$ + Na $\rightarrow$ Si + Na$^+$      & $2.7 \times 10^{-9}$                                                   & (2) \\
              45 & Si$^+$ + Fe $\rightarrow$ Si + Fe$^+$      & $1.9 \times 10^{-9}$                                                   & (2) \\
              46 & SiO$^+$ + N $\rightarrow$ Si$^+$ + NO      & $2.1 \times 10^{-10}$                                                  & (2) \\
              47 & SiO$^+$ + Fe $\rightarrow$ SiO + Fe$^+$    & $1 \times 10^{-9}$                                                     & (2) \\
              48 & SiO$^+$ + C $\rightarrow$ Si$^+$ + CO      & $1 \times 10^{-9}$                                                     & (2) \\
              49 & SiO$^+$ + O $\rightarrow$ Si$^+$ + O$_2$   & $2 \times 10^{-10}$                                                    & (2) \\
              50 & SiO$_2$ + He$^+$ $\rightarrow$ O$_2$ + Si$^+$ + He & $2 \times 10^{-9}$                                             & (2) \\
              51 & SiH$^+$ + H $\rightarrow$ Si$^+$ + H$_2$   & $1.9 \times 10^{-9}$                                                   & (2) \\
              52 & Si$^+$ + e$^-$ $\rightarrow$ Si + $\gamma$ & $4.26 \times 10^{-12} (\mathrm{T}/300)^{-0.61}$                        & (2) \\
              53 & SiO$^+$ + e$^-$ $\rightarrow$ Si + O       & $2 \times 10^{-7} (\mathrm{T}/300)^{-0.5}$                             & (2) \\
              54 & SiH$^+$ + e$^-$ $\rightarrow$ Si + H       & $2 \times 10^{-7} (\mathrm{T}/300)^{-0.5}$                             & (2) \\
              55 & SiF$^+$ + e$^-$ $\rightarrow$ Si + F       & $2 \times 10^{-7} (\mathrm{T}/300)^{-0.5}$                             & (2) \\
             \hline \hline
        \end{tabular}
        \tablefoot{The first block describes the Ti chemistry; the second block describes the Si chemistry. For the definition of EQR$_i$ see \citet{boulangier_developing_2019}, where $i$ is the reaction number as defined in their work. References: (0) This work, (1) \citet{campbell_kinetics_1993}, (2) \citet{mcelroy_umist_2013}, (3) \citet{plane_nucleation_2013}, (4) \citet{ritter_kinetics_2002}, (5) \citet{higuchi_kinetics_2008}, (6) \citet{kiefer_effect_2023}.}
    \end{table*}

    \onecolumn
    \LTcapwidth=\textwidth
    \begin{longtable}{l l l l l l l}
        \caption{\label{tab:con_reac} Bulk growth reactions considered for this study. The list of reaction was taken from \citet{helling_sparkling_2019}. s$_0$, s$_1$, s$_2$ and s$_3$ are the fitting parameters of the vapor number density (Eq. \ref{eq:s_fit}). $\Delta$V is the volume increase per bulk growth reaction.} \\
        \hline\hline
        nr & reaction & s$_0$ & s$_1$ [K] & s$_2$ [K$^2$] & s$_3$ [K$^3$] & $\Delta$V [cm$^{-3}$] \\
        \hline
        \endfirsthead
        \caption{continued.}\\
        \hline\hline
        nr & reaction & s$_0$ & s$_1$ [K] & s$_2$ [K$^2$] & s$_3$ [K$^3$] & $\Delta$V [cm$^{-3}$] \\
        \hline
        \endhead
        \hline
        \endfoot
            c1 & Ti + 2 H$_2$O $\rightarrow$ TiO$_2$[s] + 2 H$_2$ & 7.902e+01 & -1.622e+05 & -4.690e+06 & 9.013e+08 & 3.136e-23 \\
            c2 & TiO$_2$ $\rightarrow$ TiO$_2$[s] & 6.194e+01 & -6.731e+04 & -5.024e+06 & 9.890e+08 & 3.136e-23 \\
            c3 & TiO + H$_2$O $\rightarrow$ TiO$_2$[s] + H$_2$ & 6.431e+01 & -7.986e+04 & -5.635e+06 & 1.084e+09 & 3.136e-23 \\
            c4 & TiS + 2 H$_2$O $\rightarrow$ TiO$_2$[s] + H$_2$S + H$_2$ & 5.244e+01 & -2.031e+05 & -4.500e+05 & 1.000e+00 & 3.136e-23 \\
            \hline
            c5 & 2 Mg + SiO + 3 H$_2$O $\rightarrow$ Mg$_2$SiO$_4$[s] + 3 H$_2$ & 1.695e+02 & -2.951e+05 & -1.166e+07 & 2.383e+09 & 7.278e-23 \\
            c6 & 2 MgOH + SiO + H$_2$O $\rightarrow$ Mg$_2$SiO$_4$[s] + 2 H$_2$ & 1.429e+02 & -1.986e+05 & -1.275e+07 & 2.560e+09 & 7.278e-23 \\
            c7 & 2 Mg(OH)$_2$ + SiO $\rightarrow$ Mg$_2$SiO$_4$[s] + H$_2$O + H$_2$ & 8.983e+01 & -3.337e+05 & -4.754e+06 & 1.094e+09 & 7.278e-23 \\
            c8 & 2 MgH + SiO + 3 H$_2$O $\rightarrow$ Mg$_2$SiO$_4$[s] + 4 H$_2$ & 1.591e+02 & -2.695e+05 & -1.247e+07 & 2.518e+09 & 7.278e-23 \\
            c9 & 2 Mg + SiS + 4 H$_2$O  & & & & & \\
               & $\rightarrow$ Mg$_2$SiO$_4$[s] + H$_2$S + 3 H$_2$ & 1.590e+02 & -4.198e+05 & -1.889e+06 & 1.000e+00 & 7.278e-23 \\
            c10 & 2 MgOH + SiS + 2 H$_2$O  & & & & & \\
               & $\rightarrow$ Mg$_2$SiO$_4$[s] + H$_2$S + 2 H$_2$ & 1.304e+02 & -3.169e+05 & -9.041e+06 & 1.909e+09 & 7.278e-23 \\
            c11 & 2 Mg(OH)$_2$ + SiS  & & & & & \\
                & $\rightarrow$ Mg$_2$SiO$_4$[s] + H$_2$ + H$_2$S & 8.892e+01 & -3.312e+05 & -5.254e+06 & 1.201e+09 & 7.278e-23 \\
            c12 & 2 MgH + SiS + 4 H$_2$O  & & & & & \\
                & $\rightarrow$ Mg$_2$SiO$_4$[s] + H$_2$S + 4 H$_2$ & 1.489e+02 & -3.948e+05 & -2.229e+06 & 1.000e+00 & 7.278e-23 \\
            c13 & 4 Mg + 2 SiH + 8 H$_2$O $\rightarrow$ 2 Mg$_2$SiO$_4$[s] + 9 H$_2$ & 1.731e+02 & -3.558e+05 & -1.181e+07 & 2.406e+09 & 7.278e-23 \\
            c14 & 4 MgOH + 2 SiH + 4 H$_2$O  & & & & & \\
                & $\rightarrow$ 2 Mg$_2$SiO$_4$[s] + 7 H$_2$ & 1.466e+02 & -2.594e+05 & -1.290e+07 & 2.583e+09 & 7.278e-23 \\
            c15 & 4 Mg(OH)$_2$ + 2 SiH $\rightarrow$ 2 Mg$_2$SiO$_4$[s] + 5 H$_2$ & 1.051e+02 & -2.737e+05 & -9.116e+06 & 1.875e+09 & 7.278e-23 \\
            c16 & 4 MgH + 2 SiH + 8 H$_2$O  & & & & & \\
                & $\rightarrow$ 2 Mg$_2$SiO$_4$[s] + 11 H$_2$ & 1.628e+02 & -3.303e+05 & -1.262e+07 & 2.541e+09 & 7.278e-23 \\
            \hline
            c17 & SiO + H$_2$O $\rightarrow$ SiO$_2$[s] + H$_2$ & 6.335e+01 & -5.758e+04 & -5.536e+06 & 1.075e+09 & 3.768e-23\\
            c18 & SiS + 2 H$_2$O $\rightarrow$ SiO$_2$[s] + H$_2$ + H$_2$S & 5.085e+01 & -1.759e+05 & -1.823e+06 & 4.249e+08 & 3.768e-23\\
            c19 & SiH + 2 H$_2$O $\rightarrow$ SiO$_2$[s] + 2 H$_2$ + H & 2.547e+01 & -1.214e+05 & -4.299e+06 & 8.301e+08 & 3.768e-23\\
            \hline
            c20 & 2 AlOH + H$_2$O $\rightarrow$ Al$_2$O$_3$[s] + 2 H$_2$ & 8.592e+01 & -1.837e+05 & -2.900e+06 & 1.000e+00 & 4.265e-23\\
            c21 & 2 AlH + 3 H$_2$O $\rightarrow$ Al$_2$O$_3$[s] + 4 H$_2$ & 1.015e+02 & -2.605e+05 & -9.682e+06 & 2.040e+09 & 4.265e-23\\
            c22 & Al$_2$O + 2 H$_2$O $\rightarrow$ Al$_2$O$_3$[s] + 2 H$_2$ & 8.493e+01 & -1.698e+05 & -9.002e+06 & 1.914e+09 & 4.265e-23\\
            c23 & 2 AlO2H $\rightarrow$ Al$_2$O$_3$[s] + H$_2$O & 7.369e+01 & -2.994e+05 & -7.691e+05 & 1.000e+00 & 4.265e-23\\
            \hline
            c24 & FeO $\rightarrow$ FeO[s] & 5.784e+01 & -5.214e+04 & -5.698e+06 & 1.137e+09 & 1.992e-23\\
            c25 & Fe + H$_2$O $\rightarrow$ FeO[s] + H$_2$ & 5.687e+01 & -4.231e+04 & -5.829e+06 & 1.140e+09 & 1.992e-23\\
            c26 & FeS + H$_2$O $\rightarrow$ FeO[s] + H$_2$S & 5.660e+01 & -3.904e+04 & -6.238e+06 & 1.248e+09 & 1.992e-23\\
            c27 & Fe(OH)$_2$ $\rightarrow$ FeO[s] + H$_2$O & 2.854e+00 & -2.023e+04 & -1.184e+06 & 2.648e+08 & 1.992e-23\\
            c28 & 2 FeH + 2 H$_2$O $\rightarrow$ 2 FeO[s] +3 H$_2$ & 4.592e+01 & -2.430e+04 & -5.662e+06 & 1.071e+09 & 1.992e-23\\
            \hline
            c29 & FeS $\rightarrow$ FeS[s] & 5.647e+01 & -4.348e+04 & -6.716e+06 & 1.289e+09 & 3.022e-23\\
            c30 & Fe + H$_2$S $\rightarrow$ FeS[s] + H$_2$ & 5.673e+01 & -4.675e+04 & -6.307e+06 & 1.181e+09 & 3.022e-23\\
            c31 & FeO + H$_2$S $\rightarrow$ FeS[s] + H$_2$O & 5.771e+01 & -5.659e+04 & -6.176e+06 & 1.177e+09 & 3.022e-23\\
            c32 & 2 FeH + 2 H$_2$S $\rightarrow$ FeS[s] + 3 H$_2$ & 4.579e+01 & -2.874e+04 & -6.140e+06 & 1.111e+09 & 3.022e-23\\
            \hline
            c33 & 2 Fe + 3 H$_2$O $\rightarrow$ Fe$_2$O$_3$[s] + 3 H$_2$ & 1.097e+02 & -2.119e+05 & -9.578e+06 & 1.885e+09 & 5.032e-23\\
            c34 & 2 FeO + H$_2$O $\rightarrow$ Fe$_2$O$_3$[s] + H$_2$ & 9.543e+01 & -1.597e+05 & -1.067e+07 & 2.079e+09 & 5.032e-23\\
            c35 & 2 Fe(OH)$_2$ $\rightarrow$ Fe$_2$O$_3$[s] + H$_2$O + H$_2$ & 2.885e+01 & -2.487e+05 & -1.944e+06 & 4.502e+08 & 5.032e-23\\
            c36 & 2 FeS + 3 H$_2$O $\rightarrow$ Fe$_2$O$_3$[s] + 2 H$_2$S + H$_2$ & 8.260e+01 & -2.675e+05 & -6.998e+06 & 1.433e+09 & 5.032e-23\\
            c37 & 2 FeH + 3 H$_2$O $\rightarrow$ Fe$_2$O$_3$[s] + 4 H$_2$ & 9.879e+01 & -1.939e+05 & -9.411e+06 & 1.816e+09 & 5.032e-23\\
            \hline
            c38 & Mg + H$_2$O $\rightarrow$ MgO[s] + H$_2$ & 5.870e+01 & -5.209e+04 & -4.378e+06 & 8.392e+08 & 1.869e-23\\
            c39 & 2 MgOH $\rightarrow$ 2 MgO[s] + H$_2$ & 2.061e+01 & -7.653e+04 & -1.259e+06 & 2.587e+08 & 1.869e-23\\
            c40 & Mg(OH)$_2$ $\rightarrow$ MgO[s] + H$_2$O & 5.951e+00 & -3.197e+04 & -4.611e+05 & 1.100e+08 & 1.869e-23\\
            c41 & 2 MgH + 2 H$_2$O $\rightarrow$ 2 MgO[s] + 3 H$_2$ & 4.839e+01 & -2.657e+04 & -5.190e+06 & 9.741e+08 & 1.869e-23\\
            \hline
            c42 & SiO $\rightarrow$ SiO[s] & 6.025e+01 & -4.653e+04 & -1.387e+06 & 2.685e+08 & 3.358e-23\\
            c43 & SiS + H$_2$O $\rightarrow$ SiO[s] + H$_2$S & 5.935e+01 & -4.397e+04 & -1.887e+06 & 3.753e+08 & 3.358e-23\\
            c44 & 2 SiH + 2 H$_2$O $\rightarrow$ 2 SiO[s] + 3 H$_2$ & 4.865e+01 & -4.527e+04 & -2.760e+06 & 4.893e+08 & 3.358e-23\\
            \hline
            c45 & 2 Fe + SiO + 3 H$_2$O $\rightarrow$ Fe$_2$SiO$_4$[s] + 3 H$_2$ & 1.679e+02 & -2.708e+05 & -1.494e+07 & 3.135e+09 & 7.708e-23\\
            c46 & 2 Fe + SiS + 4 H$_2$O  & & & & & \\
                 & $\rightarrow$ Fe$_2$SiO$_4$[s] + H$_2$S + 3 H$_2$ & 1.584e+02 & -3.983e+05 & -2.538e+06 & 1.000e+00 & 7.708e-23\\
            c47 & 4 Fe + 2 SiH + 8 H$_2$O $\rightarrow$ 2 Fe$_2$SiO$_4$[s] + 9 H$_2$ & 1.716e+02 & -3.316e+05 & -1.509e+07 & 3.158e+09 & 7.708e-23\\
            c48 & 2 FeO + SiO + H$_2$O $\rightarrow$ Fe$_2$SiO$_4$[s] + H$_2$ & 1.536e+02 & -2.186e+05 & -1.603e+07 & 3.329e+09 & 7.708e-23\\
            c49 & 2 FeO + SiS + 2 H$_2$O  & & & & & \\
                 & $\rightarrow$ Fe$_2$SiO$_4$[s] + H$_2$S + H$_2$ & 1.411e+02 & -3.369e+05 & -1.232e+07 & 2.678e+09 & 7.708e-23\\
            c50 & 4 FeO + 2 SiH + 4 H$_2$O  & & & & & \\
                 & $\rightarrow$ 2 Fe$_2$SiO$_4$[s] + 5 H$_2$ & 1.573e+02 & -2.794e+05 & -1.618e+07 & 3.352e+09 & 7.708e-23\\
            c51 & 2 FeS + SiO + 3 H$_2$O $\rightarrow$ Fe$_2$SiO$_4$[s] + 3 H$_2$S & 1.440e+02 & -3.363e+05 & -2.978e+06 & 1.000e+00 & 7.708e-23\\
            c52 & 2 FeS + SiS + 4 H$_2$O  & & & & & \\
                 & $\rightarrow$ Fe$_2$SiO$_4$[s] + 3 H$_2$S + H$_2$ & 1.416e+02 & -3.480e+05 & -1.231e+07 & 2.678e+09 & 7.708e-23\\
            c53 & 4 FeS + 2 SiH + 8 H$_2$O  & & & & & \\
                 & $\rightarrow$ 2 Fe$_2$SiO$_4$[s] + 4 H$_2$S + 5 H$_2$ & 1.477e+02 & -3.972e+05 & -3.045e+06 & 1.000e+00 & 7.708e-23\\
            c54 & 2 Fe(OH)$_2$ + SiO $\rightarrow$ Fe$_2$SiO$_4$[s] + H$_2$O + H$_2$ & 8.704e+01 & -3.075e+05 & -7.303e+06 & 1.700e+09 & 7.708e-23\\
            c55 & 2 Fe(OH)$_2$ + SiS $\rightarrow$ Fe$_2$SiO$_4$[s] + H$_2$S + H$_2$ & 8.614e+01 & -3.049e+05 & -7.803e+06 & 1.806e+09 & 7.708e-23\\
            c56 & 4 Fe(OH)$_2$ + 2 SiH $\rightarrow$ 2 Fe$_2$SiO$_4$[s] + 5 H$_2$ & 1.023e+02 & -2.474e+05 & -1.167e+07 & 2.480e+09 & 7.708e-23\\
            c57 & 2 FeH + SiO + 3 H$_2$O $\rightarrow$ Fe$_2$SiO$_4$[s] + 4 H$_2$ & 1.570e+02 & -2.528e+05 & -1.477e+07 & 3.065e+09 & 7.708e-23\\
            c58 & 2 FeH + SiS + 4 H$_2$O  & & & & & \\
                 & $\rightarrow$ Fe$_2$SiO$_4$[s] + H$_2$S + 4 H$_2$ & 1.474e+02 & -3.801e+05 & -2.614e+06 & 1.000e+00 & 7.708e-23\\
            c59 & 4 FeH + 2 SiH + 8 H$_2$O  & & & & & \\
                 & $\rightarrow$ 2 Fe$_2$SiO$_4$[s] + 11 H$_2$ & 1.607e+02 & -3.136e+05 & -1.492e+07 & 3.089e+09 & 7.708e-23\\
    \end{longtable}

    \section{Additional Plots}

\end{appendix}

\end{document}